# Tuning the mechanical behaviour of additively manufactured metamaterials with twinning and meta-harmonics

*David McArthur[1], PJ Tan[1], Chu Lun Alex Leung[1, 2, *]*

**Affiliations:**

[1] Department of Mechanical Engineering, University College London, United Kingdom

[2] Research Complex at Harwell, Harwell Campus, Didcot OX11 0FA, United Kingdom

**Authors: j.mcarthur@ucl.ac.uk  pj.tan@ucl.ac.uk ; alex.leung@ucl.ac.uk***

**Abstract:** Body-Centred Cubic (BCC) lattices with twinned meta-crystal architecture inspired by the strengthening of bulk metals have significantly improved mechanical performance; however, their deformation behaviour and underlying strengthening mechanisms remain unclear. Here, we reveal that twinning causes a transition from bending to stretch-dominated behaviour in BCC lattices, eliciting vast improvements in stiffness (+162%) and strength (+95%) without changing nodal connectivity. We designed meta-harmonic lattices by controlling a heterogenous distribution of twinned grain boundaries, inspired by bimodal harmonic microstructure, and we amplified the axial strain energy at location specific sites to further enhance the stiffness of BCC lattices by 206%. Our lattice design philosophy unleashes the potential of cellular materials for high-performance engineering applications.

**One-Sentence Summary:** Additively manufactured lattices with twinned meta-crystal design show enhanced performance and tunable deformation behaviour.

## Main Text:

### Introduction

Lightweight, high-performance lattices are of great interest for a wide range of industrial applications, including heat exchangers and biomedical implants (*1*). Analogy can be made between the architecture of lattices and the atomic arrangement of metals, in which the struts and intersection of lattices are akin to atomic bonds and atomic sites, respectively. The deformation behaviour of lattices and strengthening mechanisms of meta-crystals is also analogous to that of metals. Under uniaxial compression, 'shear bands' may form as strain localises on planes of maximum shear stress, decreasing the load-carrying and energy absorption capacities of a lattice. 'Meta-crystal' lattices comprise a new class of metamaterial with geometric features that mimic metallic microstructures, such as grain boundaries, precipitates and solutes (*2–5*), which impede and deflect the propagation of shear bands as microstructural features impede dislocation slip in metallic systems (*2*, *4*).

Reducing the meta-grain size enhances the yield strength, $\sigma_y$, of these meta-crystals following the Hall-Petch equation given by

$$\sigma_y = \sigma_0 + \frac{k}{\sqrt{d_{eq}}} \qquad [2]$$

where $\sigma_0$ is the frictional stress, $k$ is a material constant and $d_{eq}$ is the equivalent grain size. Pham *et al.* (*2*) introduced a cubic frame between meta-grains in FCC lattices to which open-ended struts are connected, increasing local relative density and altering nodal connectivity at the meta-grain boundaries. The resulting impediment of shear band propagation might be caused by variations in local relative density at the meta-grain boundaries, the reconnected struts, and the cubic frame itself. Bian *et al*. (*6*) introduced twinned meta-crystals to BCC lattices whilst maintaining their nodal connectivity and local relative density but did not identify the fundamental strengthening mechanism or test the design experimentally. Heterogenous lattice architectures, such as Functionally Graded Materials (FGM), meta-precipitates and disordered metamaterials have also been reported to improve strength and energy absorption through programmable strain localisation and fracture behaviour (*2*, *7*, *8*).

Here, we offer a new twinning strategy (depicted in **Figure 1A** and **supplementary materials**) that allows local control of deformation mode, stiffening and strengthening of heterogenous metamaterials using Twinned BCC (BCCT) lattice architecture inspired by bimodal harmonic microstructure (*9*). At the strut scale, the deformation response of lattices is largely determined by their nodal connectivity and can be classified as either bending-dominated (low nodal connectivity) or stretch-dominated (high nodal connectivity) (*10*). Stretch-dominated lattices typically exhibit higher stiffness and strength compared to their bending-dominated counterparts and thus, control of the dominant deformation mode grants engineers control of the mechanical properties. Zhao *et al.* (*11*) successfully tuned the dominant deformation mode of hybrid cubic-BCC lattices from bending to stretch but also altered the nodal connectivities by reorienting individual struts.

We have systematically designed and tested a series of BCCT lattices with different twinning angles, θ, and meta-harmonic architecture following the procedure depicted in **Figure 1A** and **supplementary materials**. These BCCT lattices have tunable deformation behaviour, both globally and locally, and controllable mechanical properties without additional weight, local density variations, or changes to nodal connectivity. We elucidate and quantify the relationships between geometric design parameters, mechanical properties and deformation behaviour of the BCCT lattices. For the first time, we demonstrate that the combination of twinning and bimodal harmonic lattice architecture can enhance the stiffness and strength of BCC lattice by 206% and 95%, respectively.

## Bulk Properties of BCC and BCCT lattices

We first predicted the bulk mechanical properties of BCCT lattices along the yy-direction, which are unaffected by free edge effects, for different twinning angles through Unit Cell (UC) analyses (see **Methods**). The stiffness of the BCCT lattices surpasses the BCC lattice and is highest for $BCCT_{45}$ ($\theta = 45°$) (see **Figure 1B** and **S-1** for Young's modulus for $27° \leq \theta \leq 45°$). Twinning also increases the yield strength of BCC lattices (**Figure 1C** and **S-1**); however, the twinning angle at which maximum yield strength occurs depends on the relative density, $\bar{\rho}$. $BCCT_{32}$ ($\theta = 32°$) and $BCCT_{37}$ ($\theta = 37°$) exhibit the highest yield strength when $\bar{\rho} \leq 0.034$ and $\bar{\rho} \geq 0.044$, respectively.

The bulk properties of the BCC lattice scale with $\bar{\rho}$ as expected of an ideal bending-dominated lattice with low nodal connectivity; $\frac{E^*}{E_s} \propto \bar{\rho}^2$ and $\frac{\sigma_y^*}{\sigma_{ys}} \propto \bar{\rho}^2$. However, the bulk properties of the BCCT lattices, with the same nodal connectivity ($M < 0$, see **Methods** sub-section '**Maxwell Criterion**'), scale with $\bar{\rho}$ as expected of an ideal stretch-dominated lattice; $\frac{E^*}{E_s} \propto \bar{\rho}$ and $\frac{\sigma_y^*}{\sigma_{ys}} \propto \bar{\rho}$. Twinning enhances bulk properties without altering the nodal connectivity and mitigates their reduction with decreasing $\bar{\rho}$. The pre-exponent and exponent values for typical scaling laws suggest that the BCCT lattice architecture causes a transition from bending to stretch-dominated behaviour and violates the G-A model (*12*). When $0.076 \leq \bar{\rho} \leq 0.3$, for which the G-A model is invalid, twinning also enhances the bulk properties through an apparent transition towards stretch-dominated behaviour; $\frac{E^*}{E_s} \propto \bar{\rho}^{2.57}$ and $\frac{\sigma_y^*}{\sigma_{ys}} \propto \bar{\rho}^{1.81}$ for the BCC lattice, $\frac{E^*}{E_s} \propto \bar{\rho}^{1.79}$ for the $BCCT_{45}$ lattice, and $\frac{\sigma_y^*}{\sigma_{ys}} \propto \bar{\rho}^{1.51}$ for the $BCCT_{37}$ lattice.

To confirm the transition in dominant deformation mode from bending to stretch, strain energy partitioning was performed (see **Methods** sub-section '**Strain Energy Partitioning**'). Strain energy increases monotonically with $\theta$ (see **Figure 1D**) and the $BCCT_{45}$ lattice has 673% higher strain energy density than the BCC, accounting for vast improvements in bulk stiffness; the $BCCT_{45}$ lattice is ~6.5 times stiffer than the BCC lattice at $\bar{\rho} = 0.14$. Enhancements in stored energy are driven by an increase in axial strain energy and hence stretch deformation. The BCC lattice is highly bending-dominated as 97.4% of the strain energy stored in the lattice derives from bending deformation. As $\theta$ increases, the proportion of axial strain energy increases while that of bending strain energy decreases; twinning causes a transition from bending- to stretch-dominated behaviour in the BCCT lattices. A mixed-mode deformation exists for $\theta = 21°$, at which strain energy derives roughly evenly from axial and bending deformation ($\eta_{axial} = 41.1\%$ and $\eta_{bending} = 56.2\%$). For $27° \leq \theta \leq 45°$, lattice deformation converges towards stretch-dominated with $68.8\% \leq \eta_{axial} \leq 95.4\%$ and $4.4\% \leq \eta_{bending} \leq 29.8\%$. This contradicts the claim by Bian *et al.* (*6*) that $BCCT_{27}$ lattices are bending-dominated. Zhao *et al.* (*11*) reoriented struts of hybrid cubic-BCC lattices and tuned the dominant deformation mode by altering nodal connectivity. However, for the first time, our results demonstrate that the dominant deformation mode of BCCT lattices can be tuned independently of nodal connectivity.

As θ varies, there is a redistribution of strain energy amongst different strut orientations. The BCC lattice has evenly distributed strain energy amongst all struts ($\delta_1 = \delta_2 = \delta_3 = \delta_4 = 25\%$). The proportion of strain energy stored in strut 1 increases with θ to 48.8% at $\theta = 27°$, driven by global lateral expansion imposing high tensile strains on strut 1 and an increase in its axial strain energy as it is reoriented towards the loading plane; these observations align with literature (*13*). For $\theta \geq 27°$, $\delta_4$ increases with θ from 5.0% ($\theta = 27°$) to 37.7% ($\theta = 45°$) as strut 4 undergoes greater axial compression when aligned increasingly towards the loading direction. For the bulk property relationships, viz. $\frac{E^*}{E_s} = A\bar{\rho}^a$ and $\frac{\sigma_y^*}{\sigma_{ys}} = B\bar{\rho}^b$, the change in exponents, from typical bending-dominated values ($a \approx b \approx 2$) for the BCC lattice to typical stretch-dominated values ($a \approx b \approx 1$) for the BCCT lattices ($\theta \geq 27°$) is caused by a transition to stretch-dominated behaviour as θ increases, most significantly in strut 1 for $\theta \leq 27°$ and strut 4 for $27° \leq \theta \leq 45°$. Strain energy partitioning is also presented for $\bar{\rho} = 0.076$ in **S-2** and at $1\% \leq \varepsilon \leq 3\%$ in **S-3** to capture the full elastic regime.

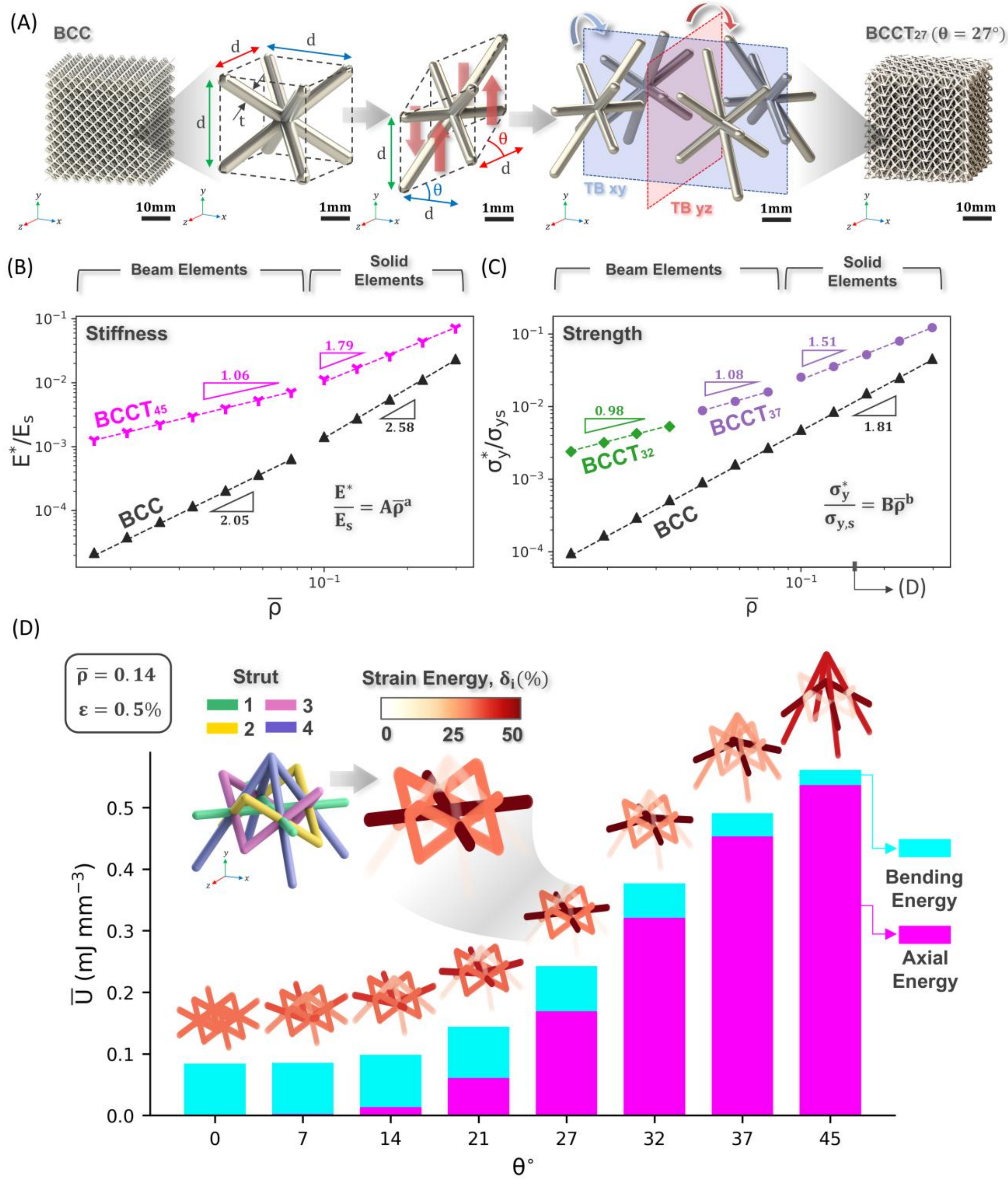

**Figure 1:** (**A**) Design process for BCCT lattices in which a BCC cell $(d \times d \times d)$ is sheared in and reflected about the xy and yz planes, tessellated in the x, y and z directions and trimmed to a cube; **(B)** Bulk relative stiffness and (**C**) strength are plotted against $\bar{\rho}$ for BCC/BCCT Unit Cells (UCs). Stiffness and strength scale linearly with $\bar{\rho}^a$ and $\bar{\rho}^b$ respectively, and exponent values are indicated for UC results and in **supplementary Table 1**; (**D**) Strain energy density, $\bar{U}$, is presented for BCC and BCCT lattices with $0° \leq \theta \leq 45°$ and $\bar{\rho} = 0.14$ at 0.5% global strain via 3D plots showing proportion of strain energy stored in each strut and axial and bending strain energy density plotted against θ

## Experiments

We designed and tested, under uniaxial compression, $10 \times 10 \times 10$ cells Rigid 4K polymer BCC and BCCT lattices printed through Stereolithography (SLA) at $\bar{\rho} = 0.14$ (see details in **Methods**). All lattices exhibit a high post-yield stress plateau that is maintained up to densification strains of $50\% \leq \varepsilon_d \leq 58\%$ (see **Figure 2A** and **S-4**). The BCC lattice exhibits barrelling and an 'X'-shaped pattern of localised strain (see **Figure 2B**), a typical mechanical response when fabricated from a ductile material (6.6% elongation at break for Rigid 4K polymer) (*14*). By contrast, the $BCCT_{27}$ lattice exhibits less severe barrelling and fractures at the nodes where struts 1 and 4 intersect (see **Figure 1D**). The $BCCT_{45}$ lattice exhibits localised deformation (see **Figure 2B**), resulting in a collapse of its global stress response. No significant defects in the SLA-printed lattices were observed through X-ray computed tomography (XCT) based on the resolvable resolution of 19.5μm. Surface roughness arising from the staircase effect, which may act as stress raisers and crack initiation sites on strut surfaces, is limited by the 50 μm layer thickness in printing. There is little variation in the local thickness along the strut length and local thickness at nodes was 103% and 139% greater than that along the struts for the BCC and $BCCT_{27}$ lattices, respectively, (see inset in **Figure 2A** and **S-5)**. Therefore, the differences in the observed deformation behaviour and resultant mechanical properties are mainly attributed to the geometric differences introduced by twinning.

Stiffness increases monotonically with θ; the $BCCT_{45}$ lattice is 162% stiffer than the BCC lattice (see **Figure 2C**). The stiffness and axial energy extracted from numerical simulations of BCCT lattices under similar conditions also increase monotonically with θ. The tremendous improvement in stiffness is caused by a comparable increase in stored axial energy. The $BCCT_{45}$ lattice, which deforms with $\bar{U}_{axial} = 0.4 \text{ mJ mm}^{-3}$ and $\bar{U}_{bending} = 0.08 \text{ mJ mm}^{-3}$ ($\eta_{axial} = 83.3\%$), is 496% stiffer in numerical simulations than its BCC counterpart, which deforms with $\bar{U}_{axial} = 0.0013 \text{ mJ mm}^{-3}$ and $\bar{U}_{bending} = 0.08 \text{ mJ mm}^{-3}$ ($\eta_{axial} = 1.6\%$). The effective mechanical properties are lower than their corresponding UC (bulk value) predictions due to specimen size effects (detailed in **S-6**) (*15*). There is a consistent ~50% knockdown in as-built stiffness compared to numerical predictions because the core of the SLA lattice undergoes less UV exposure compared to its peripheries during curing, creating a 'candy-shell' effect resulting in stiffer and stronger peripheries than the under-cured core (*16*).

The strength of BCCT lattices increases with θ for $\theta \leq 27°$; the $BCCT_{27}$ lattice is 128% stronger than the BCC lattice (see **Figure 2D**). However, at higher twinning angles ($\theta \geq 32°$), the strength decreases with θ; $BCCT_{32-45}$ lattices have 7%-33% lower strength than the $BCCT_{27}$ lattice due to the localisation of their deformation (see **Figure 2B**). Numerical models predict localised deformation in crush bands, as observed experimentally, and that yielding is preceded by local buckling of strut 4 that is increasingly aligned with the load direction as θ increases. No local buckling occurs for $\theta \leq 27°$ but it is proposed that local buckling and yielding contribute to a mixed mode of failure in as-built lattices when $\theta \geq 32°$ (see **Figure 2D**). Twinning also improves the stiffness and strength of BCC lattices at $\bar{\rho} = 0.07$, demonstrating that the transition from bending to stretch-dominated behaviour is achieved for relative densities at which the G-A model is valid (see **S-7**).

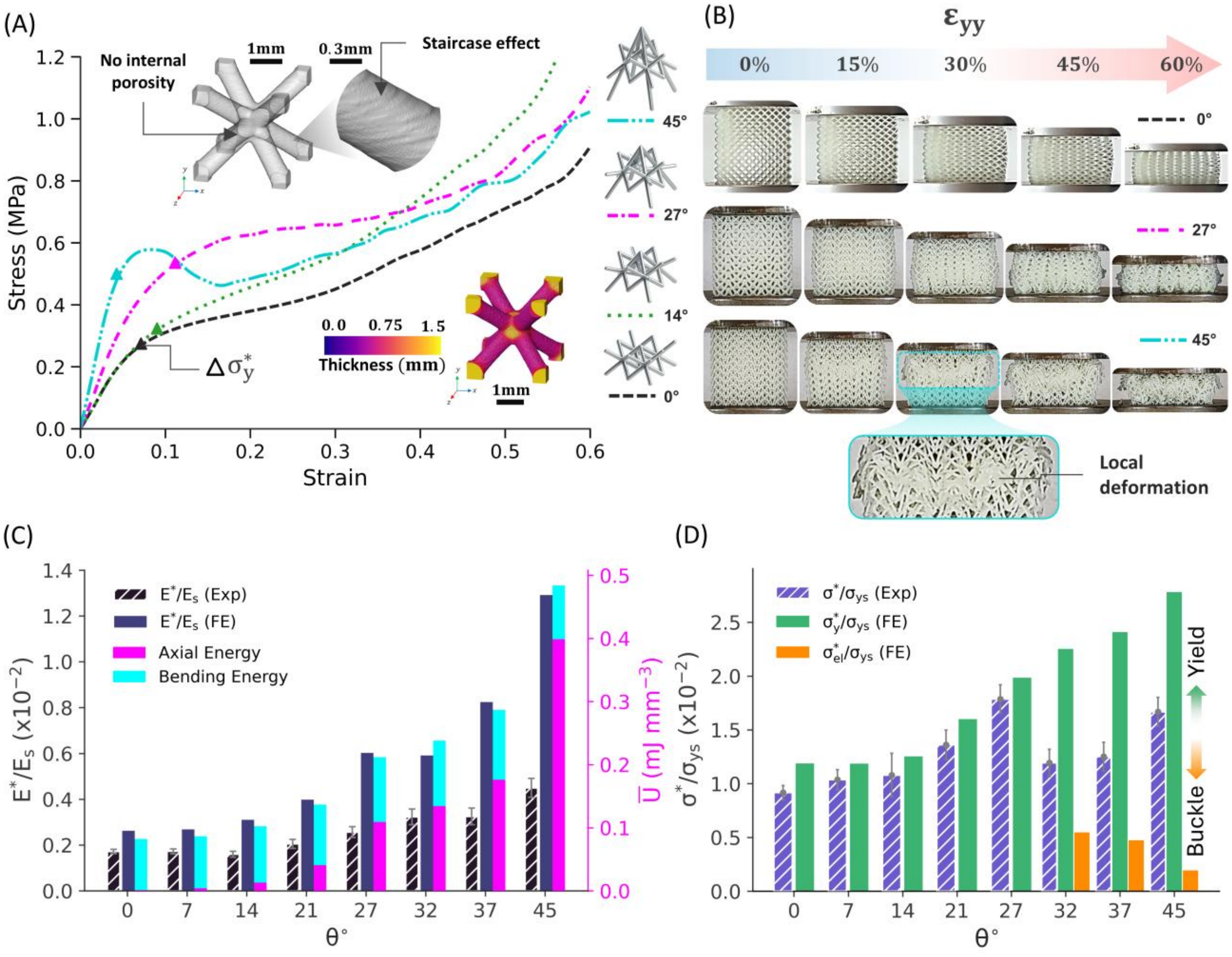

**Figure 2:** (**A**) Stress-strain response of BCCT lattices with varied θ ($\overline{\rho} = 0.14$) and (inset) 3D rendering of a BCC unit cell generated from an XCT scan, showing no internal porosity, consistent strut thickness and staircase effect from 50μm layer thickness of print; (**B**) Compression behaviour of BCCT lattices showing typical 'X'-shaped localisation in the BCC lattice, homogenous strain distribution and widespread local fractures in the $BCCT_{27}$ lattice and localised deformation in the $BCCT_{45}$ lattice; (**C**) Relative stiffness from experimental and numerical compressions and axial/bending strain energy density increase monotonically with θ for $\theta \leq 45°$; (**D**) Relative strength from experimental compressions and relative yield and buckling strength from numerical simulations plotted against θ for $\theta \leq 45°$. Relative buckling strength is not plotted for $\theta \leq 27°$ since it far exceeds yield strength.

### Meta-Grain and Meta-Harmonic Lattices

$8 \times 8 \times 8$ cell BCCT27 lattices ($\bar{\rho} = 0.14$) were designed with varied meta-grain size (see **Figure 3A, supplementary materials** and **S-8**) to investigate the effects on the effective mechanical properties. Each meta-grain has an equivalent spherical diameter, $d^{*}_{eq}$, given by

$$d^{*}_{eq} = \sqrt[3]{\frac{6\overline{G}^{2}\overline{H}}{\pi}}\,d \qquad [4]$$

where $\overline{G}$, $\overline{G}$ and $\overline{H} = 7$ are the complete number of cells in the x, z and y directions, respectively, so that the lattice has a volume-weighted mean $d_{eq}$. Meta-Grain (MG) lattices were designed with meta-grains of consistent size ($\overline{G}44$, $\overline{G}2222$ and $\overline{G}1x8$ lattices) akin to coarse and fine-grained metallic microstructures. Inspired by bimodal harmonic microstructure, that consists of coarse grains surrounded by a shell of ultra-fine grains resulting in higher effective strength (*17*), we designed Meta-Harmonic (MH) lattices with larger meta-grains surrounded by smaller meta-grains ($\overline{G}242$, $\overline{G}11411$ and $\overline{G}112211$ lattices – see **Figure 3A**). The stress-strain responses under experimental compression of MG and MH lattices printed in Rigid 4K are shown in **S-9**.

## Stiffness

The MH lattices are stiffer than the MG lattices in both experiment and numerical simulation; the as-printed $\overline{G}11411$ lattice is 27.3% stiffer than the stiffest MG lattice ($\overline{G}1x8$) (see **Figure 3B**). Numerical simulation reveals that heightened axial strain energy is responsible for such an improvement in stiffness; the $\overline{G}11411$ lattice stores 37% more axial strain energy than the $\overline{G}1x8$ lattice at $\varepsilon = 0.5\%$. 3D maps of axial strain energy reveal that, for all the MH lattices, there is greater localisation around Twin Grain Boundaries (TGBs) between differently sized meta-grains, termed 'Harmonic Grain Boundaries' (HGBs) (see **Figure 3C**). For the $\overline{G}112211$ lattice, the axial strain energy is lower at the lattice core and free surfaces and localises at the HGBs between the $\overline{G} = 1$ and $\overline{G} = 2$ meta-grains. However, the proportion of axial strain energy sharply increases at HGBs. In the $\overline{G}1x8$, the most highly stretch-dominated behaviour occurs in the lattice core, where $\eta_{axial} = 66.7\%$, but for the MH lattices, the most highly stretch-dominated regions occur at the HGBs, where $\eta_{axial} = 86.2\%$ for the $\overline{G}112211$ lattice (see **Figure 3D** and **S-10** for all MH lattices). Here, we demonstrate that the deformation mode of twinned lattices can be tailored to be more stretch- or bending-dominated, and through a distribution of smaller and larger meta-grains, the local deformation behaviour can be controlled. MH lattices with harmonic grain boundaries can be employed as another design tool to achieve the additional stiffness of meta-materials.

UC analysis reveals that for MG lattices, a reduction in $d_{eq}$ increases stiffness predominantly driven by an increase in axial strain energy and transition to more stretch-dominated behaviour (see **S-11**). However, the behaviour of 10-cell lattices differs from that of infinitely-sized lattices; the $\overline{G}44$ lattice is stiffer than the $\overline{G}222$ lattice despite its larger meta-grains. For the MG lattices, strain energy localises more severely at the intersections between TGBs and compression plates for larger meta-grains (see **S-12** and **S-13** for axial and total strain energy, respectively). Conversely, decreasing meta-grain size further to form the $\overline{G}1x8$ lattice increases the stiffness by increasing the axial strain energy in the lattice core (see **Figure 3C**). For the $\overline{\mathrm{G}}44$ and $\overline{\mathrm{G}}2222$ lattices, the dominant deformation in the lattice core is by bending as $\eta_{\mathrm{axial}} = 38.8\%$ and $38.0\%$, respectively.

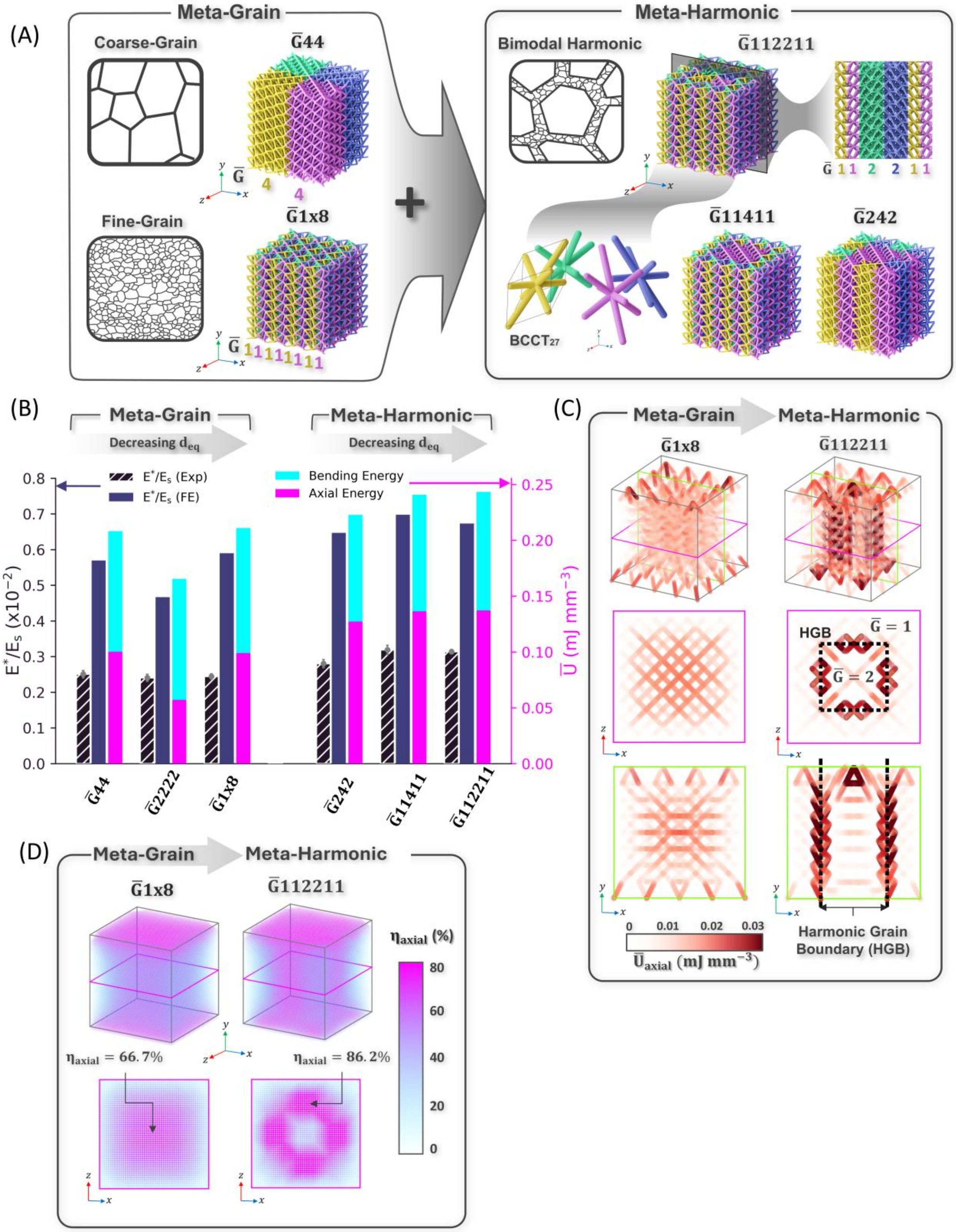

**Figure 3:** (**A**) Design of Meta-Grain (MG) and Meta-Harmonic (MH) lattices formed from differently oriented BCCT27 meta-grains of different number of complete cells $\overline{\mathrm{G}}$ in the x and z directions and $\overline{\mathrm{H}} = 7$ in the y direction. (**B**) Relative stiffness obtained from experimental compressions and numerical simulations plotted alongside axial

strain energy density, $\overline{U}_{axial}$; (C) 3D and 2D section maps plotting $\overline{U}_{axial}$ for the $\overline{G}1x8$ and $\overline{G}112211$ lattices at $\varepsilon = 0.5\%$ – Harmonic Grain Boundaries (HGBs) are indicated showing a concentration of axial strain energy at the intersection of differently sized meta-grains. (D) 3D and 2D section maps plotting $\eta_{axial}$ for $\overline{G}112211$ and $\overline{G}1x8$ lattices at a global strain, $\varepsilon = 0.5\%$

## Strength

The yield strength of MG and MH lattices increases with decreasing $d_{eq}$ in both experiment (see **Figure 4A**) and numerical simulation (see **S-14**). This correlation agrees with the effects of grain boundary strengthening and, for MG lattices, the linear relationship between yield strength and $1/\sqrt{d_{eq}}$ agrees with the Hall-Petch relationship observed in other meta-crystals (*2*, *6*, *18*). Full field 3D strain maps ($\varepsilon_{yy}$ and $\varepsilon_{xy}$) of the MG lattices (see **Figure 4B**), obtained by numerical simulations, reveal the deflection of diagonally localised shear bands at the TGBs akin to the grain boundary strengthening mechanism (*2*). $\varepsilon_{xy}$ and $\varepsilon_{yz}$ strain maps of MG lattices reveal equal and opposite shear strains for adjacent meta-grains in the yz and xy planes, and $\varepsilon_{xy} \approx \varepsilon_{yz} \approx 0$ at TGBs (see **Figure 4B** and **S-15** for all lattices). We posit that an increase in the density of TGBs increases the proportion of the lattice with negligible shear strain and reduces the shear strain within meta-grains. This effect is similar to the dislocation pile up at grain boundaries wherein plastic deformation is impeded by misoriented atomic or slip planes (*19*).

However, the MH lattices outperform the Hall-Petch predictions of strength; the MH lattices have higher strength than the MG lattices for the same $d_{eq}$. Enhancements in strength are attributed to two complementary mechanisms. Firstly, there is an enhancement in axial strain energy at HGBs causing more stretch-dominated deformation (see **Figure 3B-D**). Secondly, there is an abatement of local yielding at lattice peripheries where smaller meta-grains are introduced. For the $\overline{G}44$ & $\overline{G}2222$ (**Figure 4B**) and $\overline{G}242$ lattices (**Figure 4C**), $\varepsilon_{yy}$ strain localises at the intersection between the contact regions and the TGBs and between the contact regions and the lattice peripheries. For $\overline{G}11411$, $\overline{G}112211$ and $\overline{G}1x8$ lattices, the highly localized strain regions are occupied by $\overline{G} = 1$ meta-grains which best impede the formation of shear bands, thereby increasing the yield strength. The addition of TGBs to the peripheries of the $\overline{G}242$ lattice forms the $\overline{G}11411$ lattice and abates strain localisation as marked in **Figure 4C.** Furthermore, the $\overline{G}11411$ MH lattice, that most closely mimics the bimodal harmonic microstructure, outperforms the Hall-Petch prediction by 20.2% because while $\overline{G} = 4$ meta-grain at the lattice core increases the mean $d_{eq}$, the $\overline{G} = 1$ meta-grain at the lattice peripheries delays yield and the large meta-grain size difference at the HGBs induces high axial strain energy. Ultimately, the $\overline{G}1x8$ and $\overline{G}112211$ lattices, with the smallest $d_{eq}$ and largest concentration of TGBs for MG and MH lattices, respectively, exhibit the highest yield strength (73.5% greater than the BCC).

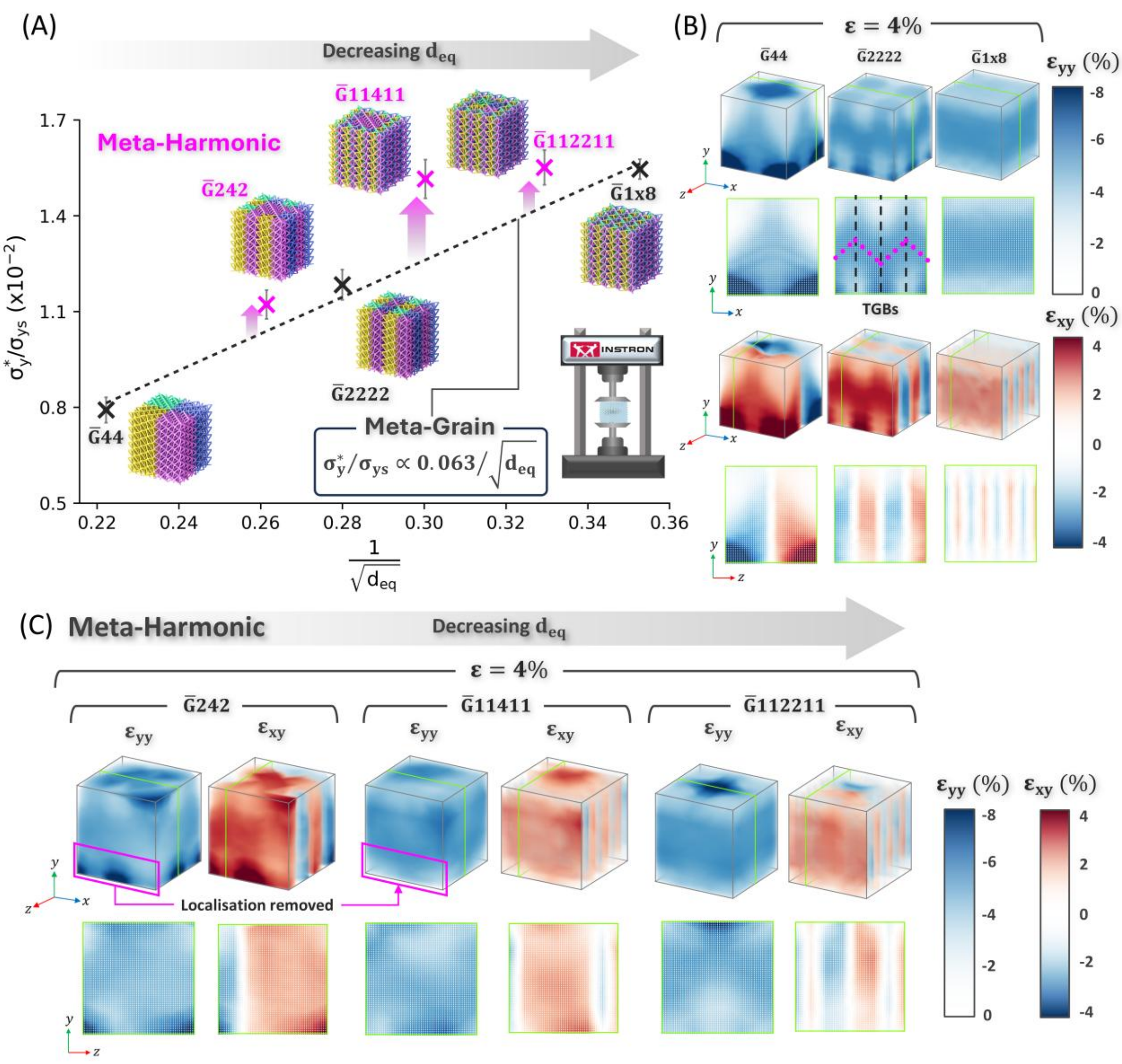

**Figure 4** (A) Relative yield strength of Meta-Grain (MG) and Meta-Harmonic (MH) lattices obtained experimentally plotted against $1/\sqrt{d_{eq}}$; a linear fit is plotted for MG lattices with $R^2 = 99\%$ showing that strengthening through meta-grains follows the Hall-Petch relationship and that MH lattices outperform this prediction of strength; (B) 3D and 2D section maps of local $\varepsilon_{yy}$ and $\varepsilon_{xy}$ strain plotted for MG lattices; (C) 3D and 2D section maps of local $\varepsilon_{yy}$ and $\varepsilon_{xy}$ strain plotted for MH lattices at global strain, $\varepsilon = 4\%$.

## Conclusions

Here, we demonstrate a series of novel design tools to tune the deformation behaviour and achieve cumulative enhancements in mechanical properties of AM BCC lattices through twinned lattice architecture with constant and variable sized meta-grains. Twinning causes a shift in deformation behaviour in BCC lattices from bending to stretch-dominated that violates the G-A model. The mechanical properties of twinned BCC lattices can be tuned to a desirable balance of stiffness and strength. We also introduce a new form of lattice architecture, Meta-Harmonic (MH) lattices with variable sized meta-grains, further enhanced axial energy, and increased stiffness beyond that of constant-sized meta-grain lattices. Axial energy localises at twin grain boundaries between meta-grains of different size, granting engineers the power to locally tune the lattice deformation behaviour. Although the yield strength of MG lattices exhibits a linear dependence on $1/\sqrt{d_{eq}}$ like the classic Hall-Petch relationship, the strength of MH lattices outperforms such predictions.

## Supplementary Materials

### Materials and Methods

#### Design

The geometry of the BCC lattice is defined by cell size, d, strut thickness, t, and number of complete cells $\overline{W}$, $\overline{H}$ and $\overline{D}$, in the x, y and z directions, respectively, given by

$$\overline{W} = W/d, \quad \overline{H} = H/d, \quad \overline{D} = D/d \qquad \text{[6], [7], [8]}$$

where W, H and D are the global dimensions in the x, y and z directions, respectively.

To form BCCT lattices, a BCC cell with coordinates $[x_1, y_1, z_1]$ is sheared along the xy and yz planes to form a BCCT lattice cell (see **S-7**) with transformed coordinates $[x_2, y_2, z_2]$ according to

$$\begin{bmatrix} x_2 \\ y_2 \\ z_2 \end{bmatrix} = \begin{bmatrix} 1 & 0 & 0 \\ \tan(\theta) & 1 & \tan(\theta) \\ 0 & 0 & 1 \end{bmatrix} \begin{bmatrix} x_1 \\ y_1 \\ z_1 \end{bmatrix} \qquad \text{[9]}$$

for which $-90° \leq \theta \leq 90°$. The resultant geometry is tessellated along vectors $\vec{r_1} = [d, d \cdot \tan(\theta), 0]$, $\vec{r_2} = [0, d \cdot \tan(\theta), d]$ and $\vec{r_3} = [0, d, 0]$ and trimmed to form a meta-grain with complete number of cells, $\overline{G}$, in the x and z directions and $\overline{H}$ in the y direction given by

$$\overline{G} = G/d, \quad \overline{H} = H/d - 1 \qquad \text{[10], [11]}$$

where G is the dimension of the meta-grain in the x and z directions and H is the dimension of the meta-grain in the y direction. Meta-grains are reflected about the xy and yz planes according to the transformations defined by

$$\begin{bmatrix} x_3 \\ y_3 \\ z_3 \end{bmatrix} = \begin{bmatrix} 2G \\ 0 \\ 0 \end{bmatrix} + \begin{bmatrix} -1 & 0 & 0 \\ 0 & 1 & 0 \\ 0 & 0 & 0 \end{bmatrix} \begin{bmatrix} x_2 \\ y_2 \\ z_2 \end{bmatrix} \qquad \text{[12]}$$

$$\begin{bmatrix} x_4 \\ y_4 \\ z_4 \end{bmatrix} = \begin{bmatrix} 0 \\ 0 \\ 2G \end{bmatrix} + \begin{bmatrix} 0 & 0 & 0 \\ 0 & 1 & 0 \\ 0 & 0 & -1 \end{bmatrix} \begin{bmatrix} x_2 \\ y_2 \\ z_2 \end{bmatrix} \qquad \text{[13]}$$

$$\begin{bmatrix} x_5 \\ y_5 \\ z_5 \end{bmatrix} = \begin{bmatrix} 2G \\ 0 \\ 2G \end{bmatrix} + \begin{bmatrix} -1 & 0 & 0 \\ 0 & 1 & 0 \\ 0 & 0 & -1 \end{bmatrix} \begin{bmatrix} x_2 \\ y_2 \\ z_2 \end{bmatrix} \qquad \text{[14]}$$

and the resultant meta-grains are assembled to form a lattice of global dimensions W, H and D and a complete number of cells $\overline{W}$, $\overline{H}$ and $\overline{D}$ in the z, y and z directions, respectively, as given by **Equation 6**, **8** and **11**, respectively (see **S-8**).

## Maxwell Criterion

At the strut scale, the deformation response can be classified as either bending-dominated or stretch-dominated (*10*), and this is largely determined by the nodal connectivity and predicted using Maxwell's stability criterion (*20*), given by

$$M = b - 3j + 6 \quad [1]$$

where b is the number of struts, j is the number of frictionless joints and M is a dimensionless parameter (*10*, *20*). In general, lattices with $M < 0$ (low connectivity) exhibit bending-dominated behaviour and those with $M \geq 0$ (high connectivity) exhibit stretch-dominated behaviour (*10*).

The Gibson-Ashby (G-A) model is a generic power-law relationship between the relative mechanical properties and the relative density of lattices (*12*). In the G-A model, the pre-exponent is governed by the lattice topology, constituent material and its fabrication method whereas the exponent, typical values of which are listed in **Table 1**, reflects the dominant deformation behaviour at the strut-scale depending on M (*10*).

**Table 1**: Scaling laws predicted by the G-A model for various relative properties of lattices

| Relative Property | Scaling Law | Ideal Exponent | Behaviour |
|---|---|---|---|
| Modulus, $\overline{E} = \frac{E^*}{E_s}$ | $\overline{E} = A\overline{\rho}^a$ | $a = 2$<br>$a = 1$ | Bending-dominated<br>Stretch-dominated |
| Yield strength, $\overline{\sigma}_y = \frac{\sigma^*}{\sigma_s}$ | $\overline{\sigma}_y = B\overline{\rho}^b$ | $\frac{3}{2} \leq b \leq 2$<br>$b = 1$ | Bending-dominated<br>Stretch-dominated |
| Elastic buckling strength, $\overline{\sigma}_{el} = \frac{\sigma^*_{el}}{E_s}$ | $\overline{\sigma}_{el} = C\overline{\rho}^c$ | $c = 2$ | Buckling-dominated |

## Materials and Fabrication

Lattices were designed in Rhinoceros 3D (McNeel, *USA*) with parametric modelling software Grasshopper (McNeel, *USA*). 3 samples of each lattice were printed with geometric parameters listed in **Table 2,** using Rigid 4K resin by SLA in a Form 3+ Low Force Stereolithography (LFS)™ 3D printer (Formlabs, *USA*). The STL models were pre-processed for printing in Preform (Formlabs, *USA*) with a minimum layer thickness of 50µm. All lattices were printed with a strut thickness of 0.7mm and the cell size, d, was varied to maintain consistent relative density of $\overline{\rho} = 0.14$. After printing, excess liquid resin was removed by Isopropyl Alcohol (IPA) washing in a Form Wash and cured for 30 minutes under 405nm light at 80℃ in a Form Cure (Formlabs, *USA*). Samples were cooled to room temperature and held in ambient conditions for a consistent length of time; ~24 hours in this study. ASTM D638 dog-bone tensile specimens were also printed by SLA using Rigid 4K resin at build angles of 0°, 45° and 90° to the build plate.

**Table 2:** Geometric parameters and measured relative densities of as-printed Rigid 4K lattices tested under compression

| Lattice architecture | $\overline{W} \times \overline{H} \times \overline{D}$ | $\overline{G}$ | d (mm) | t (mm) | $\overline{\rho}_{CAD}$ | $\overline{\rho}_{CT}$ | $\overline{\rho}_{manual}$ |
|---|---|---|---|---|---|---|---|
| BCC | $10 \times 10 \times 10$ | - | 3.96 | 0.7 | 0.14 | 0.157 | 0.160 |
| $BCCT_7$ | $10 \times 10 \times 10$ | 1x10 | 3.98 | 0.7 | 0.14 | - | 0.152 |
| $BCCT_{14}$ | $10 \times 10 \times 10$ | 1x10 | 4.00 | 0.7 | 0.14 | - | 0.154 |
| $BCCT_{21}$ | $10 \times 10 \times 10$ | 1x10 | 4.03 | 0.7 | 0.14 | - | 0.158 |
| $BCCT_{27}$ | $10 \times 10 \times 10$ | 1x10 | 4.08 | 0.7 | 0.14 | 0.152 | 0.154 |
| $BCCT_{32}$ | $10 \times 10 \times 10$ | 1x10 | 4.14 | 0.7 | 0.14 | - | 0.155 |
| $BCCT_{37}$ | $10 \times 10 \times 10$ | 1x10 | 4.22 | 0.7 | 0.14 | - | 0.152 |
| $BCCT_{45}$ | $10 \times 10 \times 10$ | 1x10 | 4.35 | 0.7 | 0.14 | - | 0.152 |
| BCC | $10 \times 10 \times 10$ | - | 5.8 | 0.7 | 0.07 | - | 0.073 |
| $BCCT_{27}$ | $10 \times 10 \times 10$ | - | 5.99 | 0.7 | 0.07 | - | 0.071 |
| $BCCT_{27}$ | $8 \times 8 \times 8$ | 44 | 4.08 | 0.7 | 0.14 | - | 0.149 |
| $BCCT_{27}$ | $8 \times 8 \times 8$ | 242 | 4.08 | 0.7 | 0.14 | - | 0.147 |
| $BCCT_{27}$ | $8 \times 8 \times 8$ | 2222 | 4.08 | 0.7 | 0.14 | - | 0.146 |
| $BCCT_{27}$ | $8 \times 8 \times 8$ | 11411 | 4.08 | 0.7 | 0.14 | - | 0.152 |
| $BCCT_{27}$ | $8 \times 8 \times 8$ | 112211 | 4.08 | 0.7 | 0.14 | - | 0.151 |
| $BCCT_{27}$ | $8 \times 8 \times 8$ | 1x8 | 4.08 | 0.7 | 0.14 | - | 0.146 |

**X-ray Computed Tomography (XCT) analysis**

Samples were scanned using an XTH 225 X-ray tomography system (Nikon, *Japan*), to quantify the local thickness distribution, internal porosity and geometric defects. Each XCT scan was performed at 150kV, 70μA and 500ms exposure time per projection for 4476 projections. The voxel size of the reconstructed volume is $23.3 \times 23.3 \times 23.3 \mu m^3$. A second scan was performed on a Region Of Interest (ROI) of each sample with a reconstructed voxel size of $6.5 \times 6.5 \times 6.5 \mu m^3$; the resolvable resolution is 19.5μm . The scans were reconstructed using beam hardening correction in XCT Pro3d (Nikon, *UK*). Post-processing of reconstructed volumes was performed in Avizo 9.0 (Thermo Fischer Scientific, *USA*). A median filter with a 3D kernel size of 26 was applied to all 16-bit grayscale images obtained from XCT scanning. An Otsu thresholding was performed to segment the lattice material and then processed for local thickness and internal porosity quantification (*21*).

**Local Thickness:** A 2D image dilation with kernel of 10 pixels was performed on the XCT scan with a voxel size of 23.3μm, followed by a 2D image erosion with a kernel of 10 pixels. The 8-bit label image stack was converted to an 8-bit binary image stack. The binarised image stack was subsequently analysed by BoneJ (McNeel, *USA*) which calculates the local thickness of struts as the diameter of the greatest sphere that can fit inside the 3D boundaries at a given point (*22*). An 8-bit grayscale image stack was produced in which pixel intensity corresponds to local thickness measurement.

**Internal Porosity:** A low Otsu threshold was applied to isolate the non-material in the sub-volume of the ROI scans. Axis connectivity was performed and the result was subtracted from the thresholded image to isolate pores completely entrained by material. The pores were separated from each other via connected component analysis and pore analysis was performed to calculate equivalent diameter ($d_{eq}$), volume, (V) and sphericity ($\psi$) of each pore following processing steps detailed by Doube *et al*. (*22*)

## Mechanical testing

Tensile testing was performed in an Instron universal testing machine at a displacement rate of $1\text{mm min}^{-1}$ and sample rate of $50\text{s}^{-1}$. Load was recorded by the Instron, and strain was recorded by an Instron strain gauge. The stress-strain response had very little dependence on build direction so an elasto-plastic material model was extracted from the mean stress-strain response for all build directions, parameters of which are presented in **Table 3**, and employed in numerical simulations. Material properties aligned well with Formlabs' manufacturer report (*23*).

**Table 3:** Material properties obtained from uniaxial tensile tests of Rigid 4K resin ASTM D638 dog-bone specimen and applied in numerical simulation

| **E (MPa)** | 3100 | **Plastic Strain, $\varepsilon_p$** | 0 | 0.0033 | 0.0068 | 0.0106 | 0.0147 | 0.0191 | 0.0237 | 0.0287 | 0.034 |
|---|---|---|---|---|---|---|---|---|---|---|---|
| **ν** | 0.33 | **Plastic Stress, $\sigma_p$ (MPa)** | 32.1 | 36.3 | 39.6 | 42.3 | 44.3 | 45.9 | 46.9 | 47.6 | 48.0 |

Quasi-static compression tests were performed in an Instron universal testing machine using a $50\text{kN}$ load cell, and the load rate, strain rate and sampling rate were $1\text{mm min}^{-1}$, $0.0245 - 0.0316\text{min}^{-1}$ and $50\text{s}^{-1}$, respectively. The recorded data was smoothed using a moving average with a window size of $1/1000\text{th}$ of the data set size. The global stress, σ and global strain, ε were calculated using

$$\sigma = \frac{F_{yy}}{A_0^G}, \qquad \varepsilon = \frac{\delta_{yy}}{H} \qquad [15], [16]$$

where $F_{yy}$ and $\delta_{yy}$ are uniaxial load and displacement, respectively, in the y-direction and global cross-sectional area $A_0^G = WD$. The energy absorption efficiency, $\Phi(\varepsilon_a)$, given by

$$\Phi(\varepsilon_a) = \frac{\int_0^{\varepsilon_a} \sigma(\varepsilon) d\varepsilon}{[\sigma(\varepsilon)]_{\varepsilon=\varepsilon_a}} \qquad [17]$$

was proposed by Tan *et al.* (*24*) as a consistent method of defining the densification strain, $\varepsilon_D$, for cellular materials under compression, $\Phi_{max} = \Phi(\varepsilon_D)$. $\Phi(\varepsilon_a)$ was used in this study to determine the densification strain and its differential, $\frac{d\Phi(\varepsilon_a)}{d\varepsilon_a}$, was used to define the limits of the elastic region.

In an idealized elastic region, $\frac{d\Phi(\varepsilon_a)}{d\varepsilon_a} = 0.5$, but in practice $\frac{d\Phi(\varepsilon_a)}{d\varepsilon_a} \approx 0.5$. The commencement of the elastic region is defined as the lowest value of ε for which $\frac{\Phi(\varepsilon_a)}{d\varepsilon_a} + T_c < 0.5$ and the end of the elastic region is defined by the yield strain, $\varepsilon_Y$, for which $\frac{d\Phi(\varepsilon_a)}{d\varepsilon_a} - T_e > 0.5$, where $T_c = T_e = 0.25$ denote the initial and final elastic thresholds, respectively. The thresholds are chosen to accommodate variations in $\frac{d\Phi(\varepsilon_a)}{d\varepsilon_a}$ not caused by contact initialisation or yielding. Having determined the start and end points of the elastic region, $E^*$ can be calculated using **Equation 18**.

$$E^* = \frac{\sigma\left(\frac{\varepsilon_Y}{2}\right) - \sigma(\varepsilon_c)}{\left(\frac{\varepsilon_Y}{2} - \varepsilon_c\right)} \qquad [18]$$

## Numerical Simulation

**Finite Lattice**

All numerical simulations were performed using Abaqus/Standard (Dassault Systèmes, *France*). Displacement controlled compressions were simulated on finite lattices modelled using B31 beam elements. Surface to node region contact was defined between 2 analytical rigid surfaces and the lattice with a friction coefficient of 0.3 between polymer and aluminium (*25*).

**Unit Cell (UC) Analysis**

UC analysis was performed in Abaqus/Standard (Dassault Systèmes, *France*) to determine the bulk properties of lattices by applying Periodic Boundary Conditions (PBCs) through in-house code and according to the methodology proposed by S. Li *et al.* (*26*), to a sub-volume of BCC and BCCT lattices (see **S-16)** for $0 \leq \theta \leq 65°$ and $0.007 \leq \bar{\rho} \leq 0.3$. UC analysis models an infinite sized lattice, free from boundary effects, from which the bulk lattice properties are obtained. Mesh convergence was performed by varying the mesh densities and obtaining the relevant mechanical properties from uniaxial compressions of UCs. The convergence criterion used is $< 3\%$ incremental change in property for an incremental increase in number of elements per strut length and per strut thickness for beam element and solid element models, respectively. Here, we used $4 - 6$ B31 beam elements per strut and $6 - 8$ C3D10 solid elements per strut (see mesh convergence in **S-17**).

B31 Timoshenko beam elements and C3D10 solid elements were used when $\bar{\rho} \leq 0.076$ and $\bar{\rho} \geq 0.1$, respectively. There must be equivalent boundary nodes on opposing surfaces of the UC. For solid elements models, this was achieved using the in-built copy mesh pattern function in Abaqus/CAE (Dassault Systèmes, *France*). Uniaxial compressions were performed by applying a displacement to a reference point to which nodal displacements were constrained.

Twinning introduces a global buckling failure mode when $\bar{\rho}$ is below the critical transition relative density, $\bar{\rho}_{crit}$, where $\bar{\rho}_{crit}$ increases with $\theta$ (see **S-1**). Buckling strength was determined through linear perturbation analysis performed using the subspace iteration solver to determine the critical buckling load as the first eigenvalue at which the model stiffness matrix becomes singular in response to a unit load applied to a reference point.

**Anisotropy**

Surface plots of Young's modulus (see **S-18**), were generated through homogenization in nTop (Dassault Systèmes, *France*) that applies 6 unit strains to a unit cell meshed with tetrahedral elements to determine the elastic constants of the anisotropic stiffness matrix given by

$$\begin{bmatrix} C_{11} & C_{12} & C_{13} & C_{14} & C_{15} & C_{16} \\ C_{21} & C_{22} & C_{23} & C_{24} & C_{25} & C_{26} \\ C_{31} & C_{32} & C_{33} & C_{34} & C_{35} & C_{36} \\ C_{41} & C_{42} & C_{43} & C_{44} & C_{45} & C_{46} \\ C_{51} & C_{52} & C_{53} & C_{54} & C_{55} & C_{56} \\ C_{61} & C_{62} & C_{63} & C_{64} & C_{65} & C_{66} \end{bmatrix} \quad [19]$$

where $C_{ij}$ refers to elastic constants in Voigt notation; $C_{ij} = \sigma_i / \varepsilon_j$. The Zener ratio, Z, is defined by

$$Z = \frac{2(C_{44} + C_{55} + C_{66})}{(C_{11} + C_{22} + C_{33}) - (C_{12} + C_{13} + C_{23})} \quad [20]$$

where $Z = 1$ for isotropic materials. Twinning reduces the anisotropy of BCC lattices; for BCCT lattices, the Zener ratio, Z, decreases monotonically with θ to $Z = 2.8$ for the BCCT45 lattice (see **S-19).**

**Strain Energy Partitioning**

The elastic strain energy of the lattice, U, is partitioned in accordance with its constituent contributions from bending $(U_{bending})$, axial stretch $(U_{axial})$, and shear $(U_{shear})$, according to Christodoulou *et al.* (*27*) and following

$$U = U_{bending} + U_{axial} + U_{shear}, \quad [21]$$

$$U_{bending} = \frac{M^2 l}{2E_s I_1}, \quad U_{axial} = \frac{F_1^2 l}{2E_s A_1} \quad \text{and} \quad U_{shear} = \frac{\psi F_2^2 l}{2G_s A_1} \quad [22], [23], [24]$$

where: M, $F_1$ and $F_2$ are bending moment, axial force and shear force, respectively, at the end nodes of each beam element, $E_s$ and $G_s$ are the Young's modulus and shear modulus, respectively, of the constituent material, $I_1$, l and $A_1$ are the second moment of area, length and cross-sectional area, respectively, of each element and ψ is a shape function ($\psi = 1.\dot{1}$ for circular cross-section struts). The total, axial, bending and shear strain energy densities: $\bar{U}$, $\bar{U}_{axial}$, $\bar{U}_{bending}$ and $\bar{U}_{shear}$, respectively, are given by

$$\bar{U} = \frac{U}{V_G}, \quad \bar{U}_{axial} = \frac{U_{axial}}{V_G}, \quad \bar{U}_{bending} = \frac{U_{bending}}{V_G}, \quad \bar{U}_{shear} = \frac{U_{shear}}{V_G} \quad [25], [26], [27], [28]$$

where $V_G$ is the global volume such that $\bar{\rho} = V^*/V_G$ where $V^*$ is the volume of lattice material. The ratios of bending and axial strain energies to total strain energy, $\eta_{bending}$ and $\eta_{axial}$, respectively, are

$$\eta_{bending} = \frac{U_{bending}}{U_{bending} + U_{axial} + U_{shear}}, \quad \eta_{axial} = \frac{U_{axial}}{U_{bending} + U_{axial} + U_{shear}} \quad [29], [30]$$

and $\bar{U}$ is partitioned into contributions, $\bar{U}_i$, for each strut numbered $i = 1, 2, 3, 4$, marked in **Figure 1E**, where the proportional contribution is $\delta_i = \frac{\bar{U}_i}{\bar{U}}$.

**Size effects**

Uniaxial compressions of BCC and BCCT27 lattices, modelled with B31 beam elements, were performed with varied number of complete cells. The results are presented in **S-6** and showed that the effective stiffness and strength of the BCC lattice does not vary significantly with lattice size but there is significant softening of the BCCT27 lattice with a reduced number of cells; the effective properties of the BCCT27 lattice, with 8-10 complete cells are ~50% of the bulk properties obtained from UC analyses.

**3D strain energy maps**

Axial, bending and shear strain energy values for each B31 beam element were extracted alongside nodal cartesian coordinates. Volume-weighted mean values of strain energies were calculated for each lattice cell. $\eta_{bending}$ and $\eta_{axial}$ were calculated for each lattice cell using **Equation 26** and **27**, respectively. 3D linear interpolation was performed using the SciPy package (SciPy, *USA*) in Python (Python Software Foundation, *USA*) to determine the strain values for a $60 \times 60 \times 60$ linearly spaced 3D point cloud within the lattice bounds. For visualization, transparency was applied to data as alpha values (0 to 1) that corresponded linearly to the data range 0 to $\eta_{axial}^{max}$.

**3D full-field strain maps**

3D full-field strain maps were generated using data from numerical simulation of finite lattices (with B31 beam elements) under uniaxial compression following steps detailed in **S-20**. Cartesian coordinates and displacements of lattice vertices were extracted from numerical simulations, see **S-20 (A-B)**. 3D Delauney triangulation was performed using the SciPy package in Python to create a tetrahedral mesh between the lattice vertices, see **S-20 (C)**. The mesh elements were treated as linear tetrahedral, see **S-20 (D)**, and a 3D strain matrix was calculated for each element, see **S-20 (E)**, following procedure detailed by Kattan *et al*. (*28*), from which volume-weighted mean strain values were calculated for each lattice cell, see **S-20 (F)**. 3D linear interpolation was performed to determine strain values for a $60 \times 60 \times 60$ linearly spaced 3D point cloud within the lattice bounds, see **S-20 (G)**. For visualization, transparency was applied to data as alpha values (0 to 1) that corresponded linearly to the data range 0 to $\varepsilon_{local}^{max}$.

## Acknowledgements

CLAL is grateful for the support from the UKRI - EPSRC grants (references: EP/R511638/1, EP/W006774/1, EP/P006566/1, EP/W003333/1, EP/V061798/1, and EP/W037483/1) and the IPG Photonics/ Royal Academy of Engineering Senior Research Fellowship in SEARCH (ref: RCSRF2324-18-71). We also acknowledge the provision of an educational license of nTop by nTopology (*Dassault Systèmes, France*).

## Author contributions

DM, PJT and CLAL conceived the project. DM and CLAL led the experimental design. DM, CLAL and PJT led the interpretation of results, and composed the manuscript. DM led the lattice design, experiments, and data analysis. DM and PJT led the numerical analysis. CLAL secured funding and project management.

# Supplementary Text

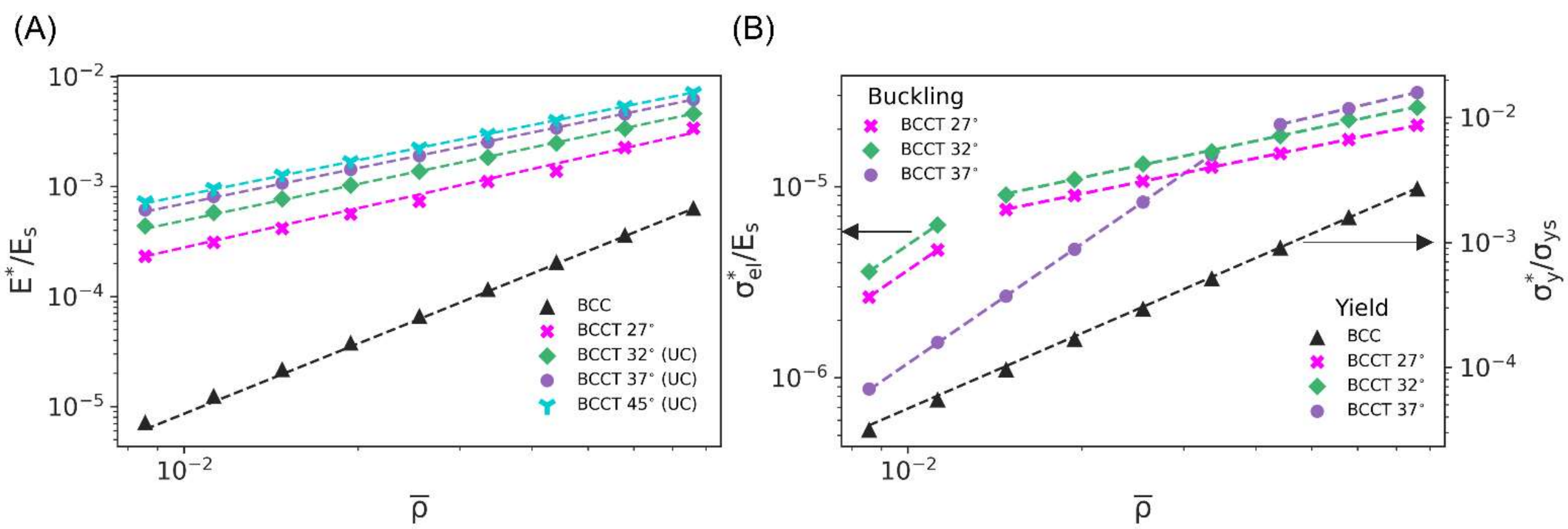

**S-1:** Bulk properties are plotted against $\overline{\rho}$ for BCCT UCs (A) stiffness for $27° \leq \theta \leq 45°$ and (B) yield strength and buckling strength, $\frac{\sigma^*_{el}}{E_s}$, for $27° \leq \theta \leq 37°$ for which strength is maximum.

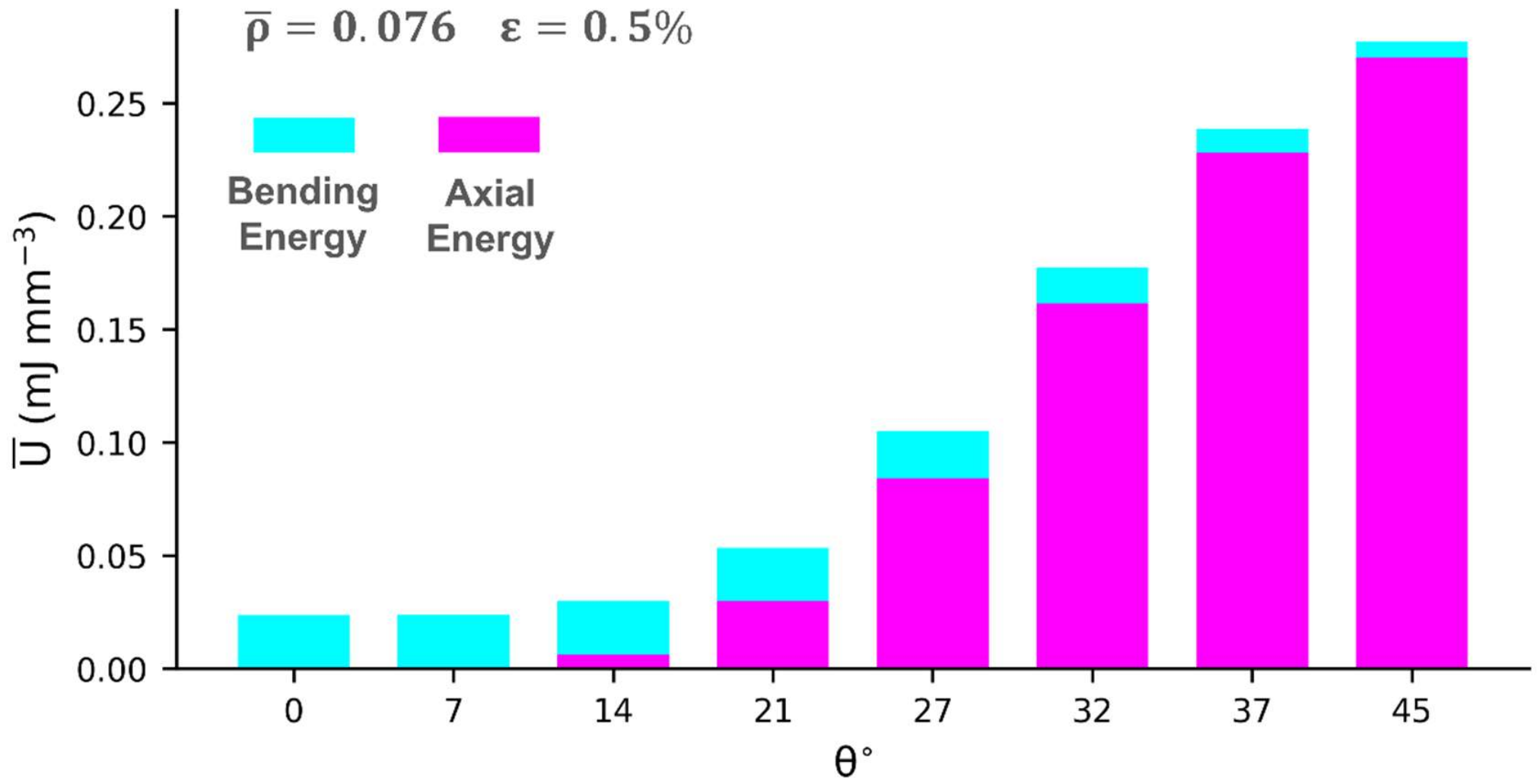

**S-2**: Strain energy density, $\overline{U}$, is presented for BCC and BCCT lattices with $0° \leq \theta \leq 45°$ and $\overline{\rho} = 0.076$ at 0.5% global strain. $\overline{U}$ is separated into contributions from axial and bending deformation showing a monotonic increase in axial deformation with $\theta$

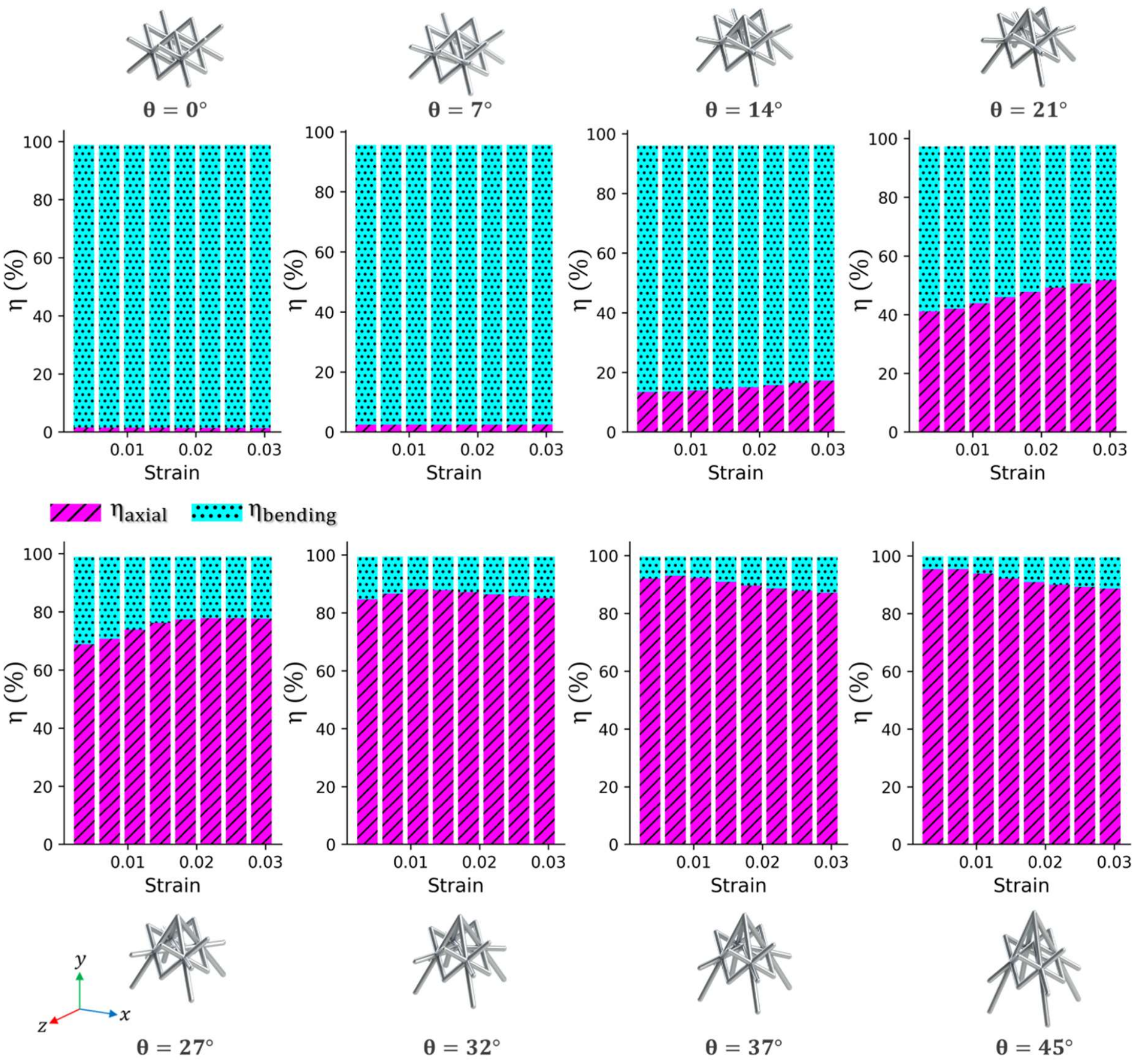

**S-3**: Proportions of strain energy, $\eta_{\mathrm{axial}}$ and $\eta_{\mathrm{bending}}$, from axial and bending deformation, respectively, for BCCT UCs with $0° \leq \theta \leq 45°$ plotted against global strain $\varepsilon \leq 3\%$ under uniaxial compression in numerical simulations

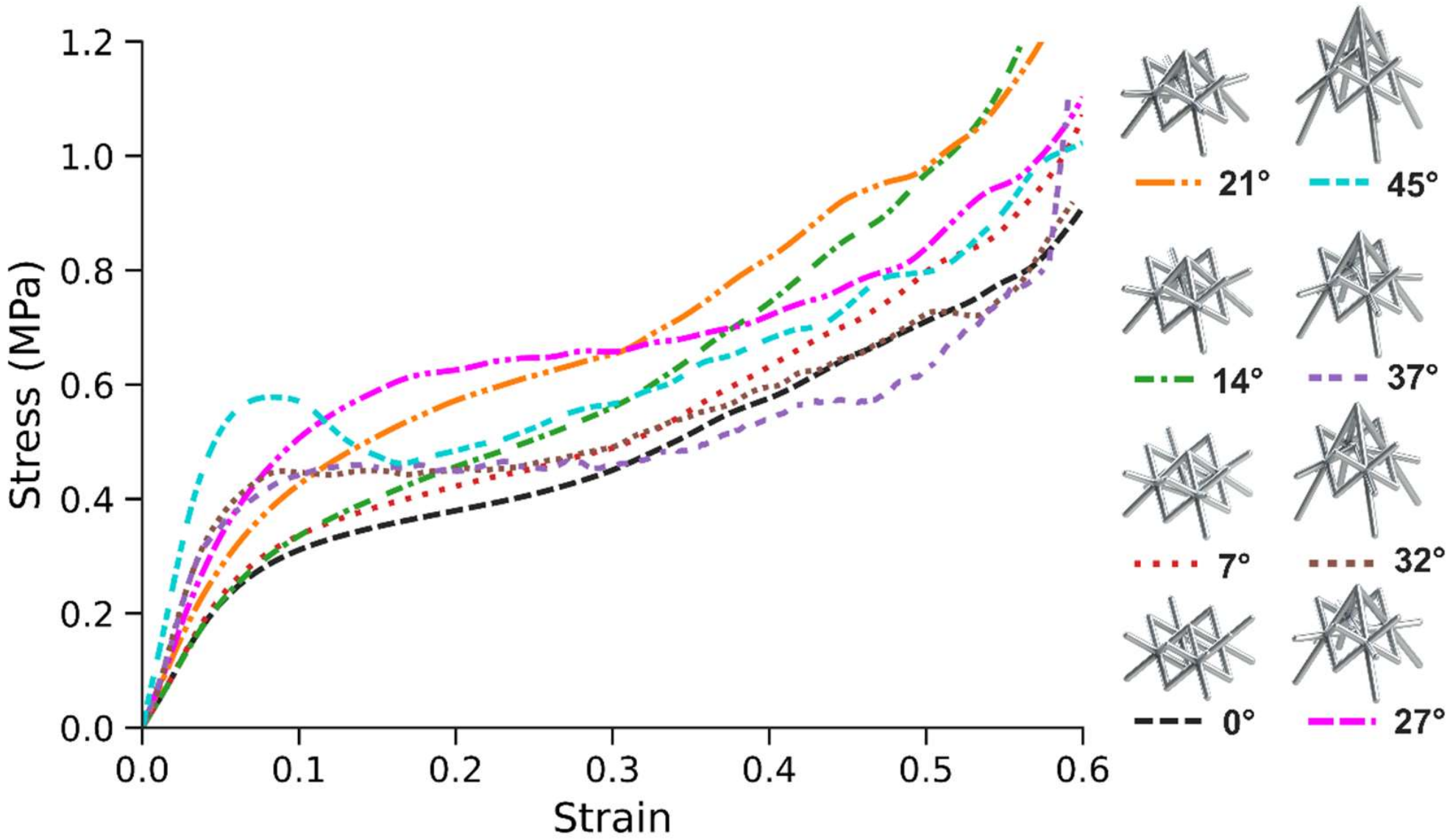

**S-4**: Stress-strain response of $10 \times 10 \times 10$ cells BCCT lattices ($\bar{\rho} = 0.14$) with $0° \leq \theta \leq 45°$ printed by SLA in Rigid 4K polymer and tested under uniaxial compression in an Instron testing machine

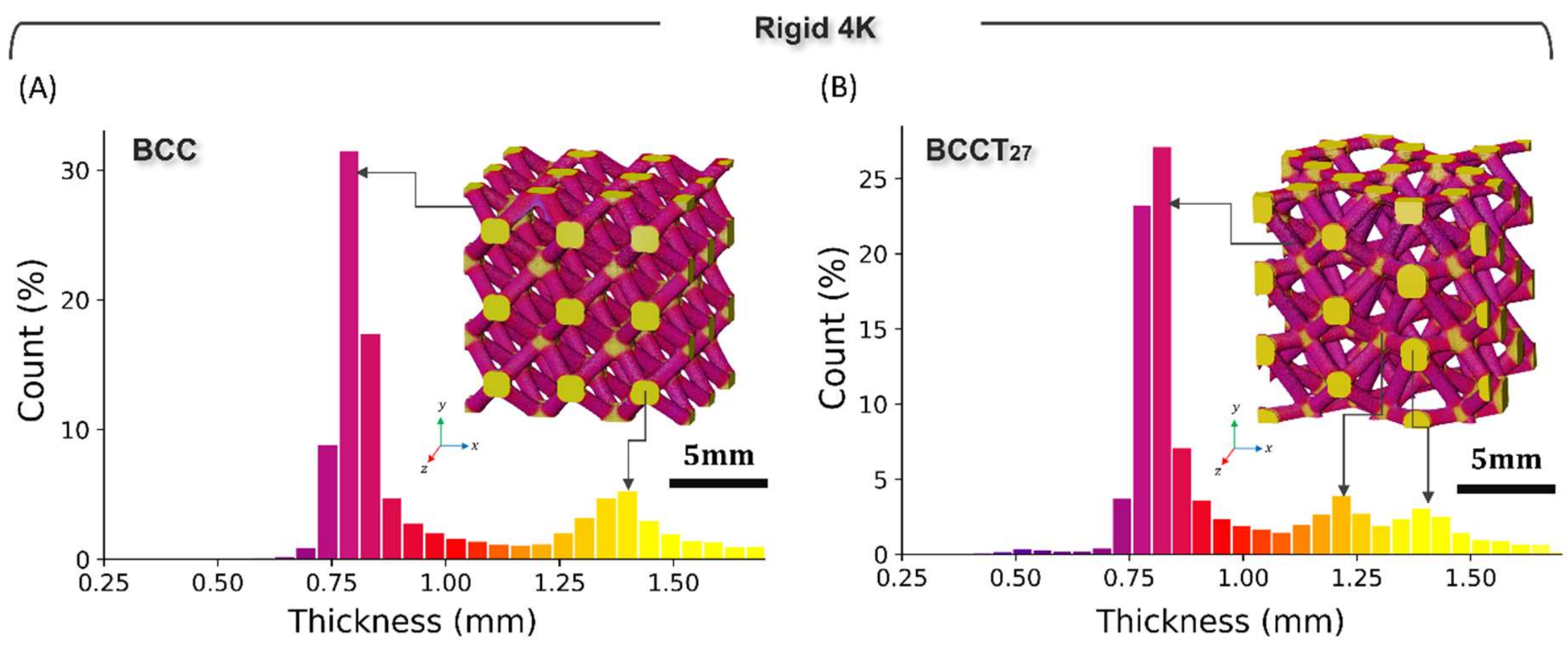

**S-5**: Local thickness histograms for $10 \times 10 \times 10$ cells (A) BCC and (B) BCCT27 lattices printed by SLA in Rigid 4K polymer with $t = 0.7$mm and $\bar{\rho} = 0.14$ and (inset) sub-volumes plotting local thickness measurements as 3D contours to visualise consistent thickness along struts and increased thickness at nodes

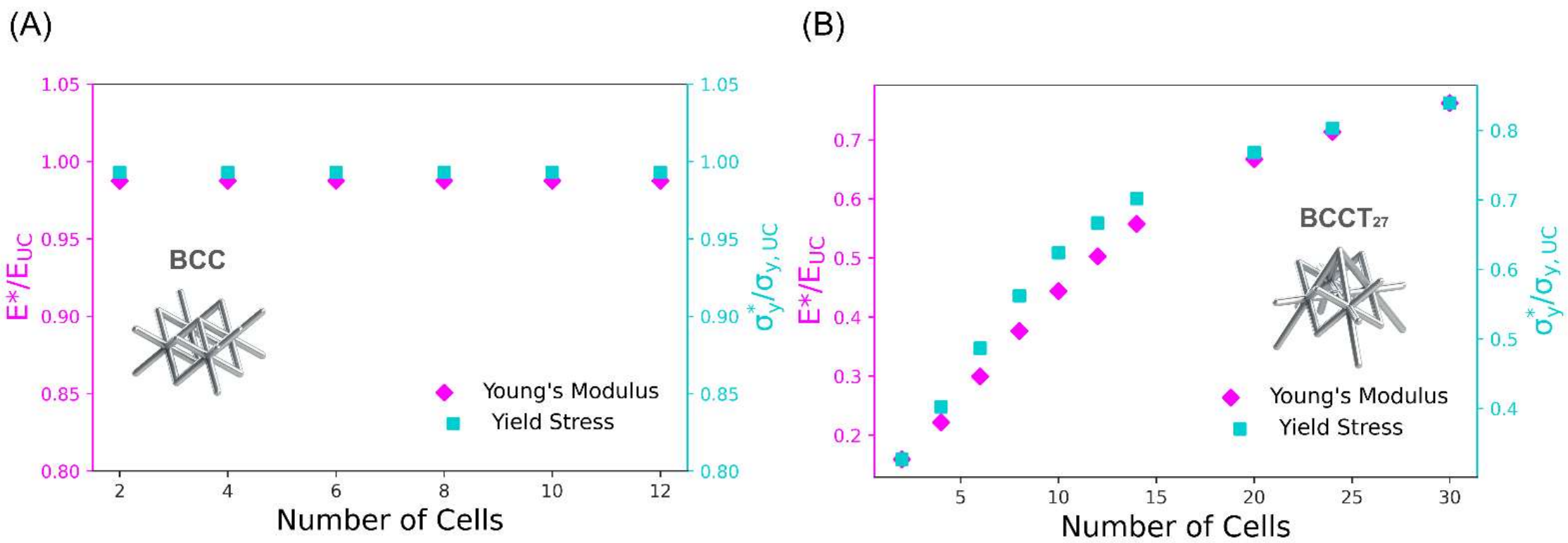

**S-6**: Effect of additional lattice cells on effective stiffness, $E^*$, and effective yield strength, $\sigma_y^*$, normalized by bulk stiffness, $E_{UC}$, and bulk strength, $\sigma_{y,UC}$, of (A) BCC lattices and (B) BCCT27 lattices tested through numerical simulation at $\bar{\rho} = 0.14$.

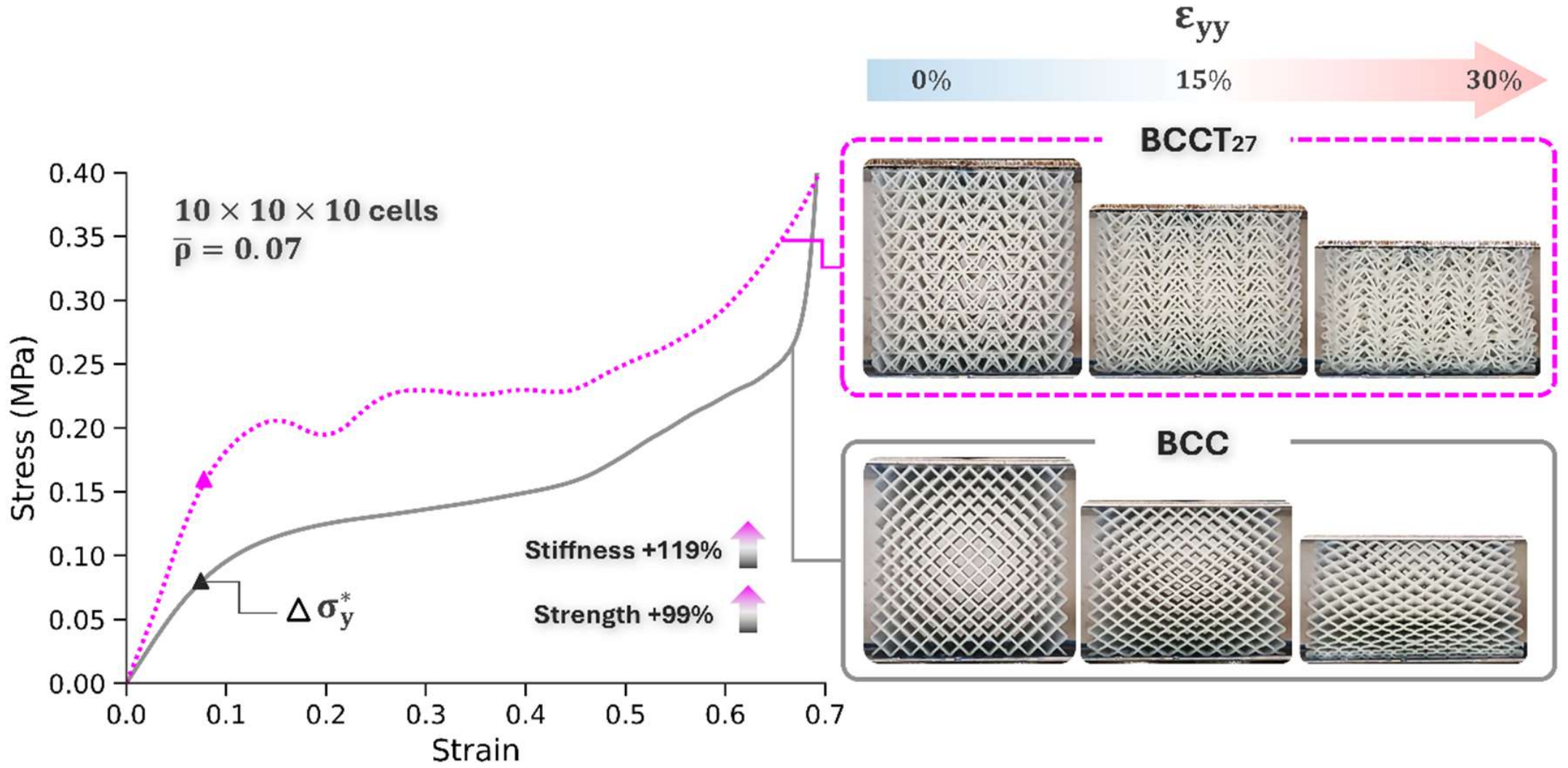

**S-7**: Stress-strain response of BCC and BCCT27 $10 \times 10 \times 10$ cell lattices printed in Rigid 4K at $\bar{\rho} = 0.07$

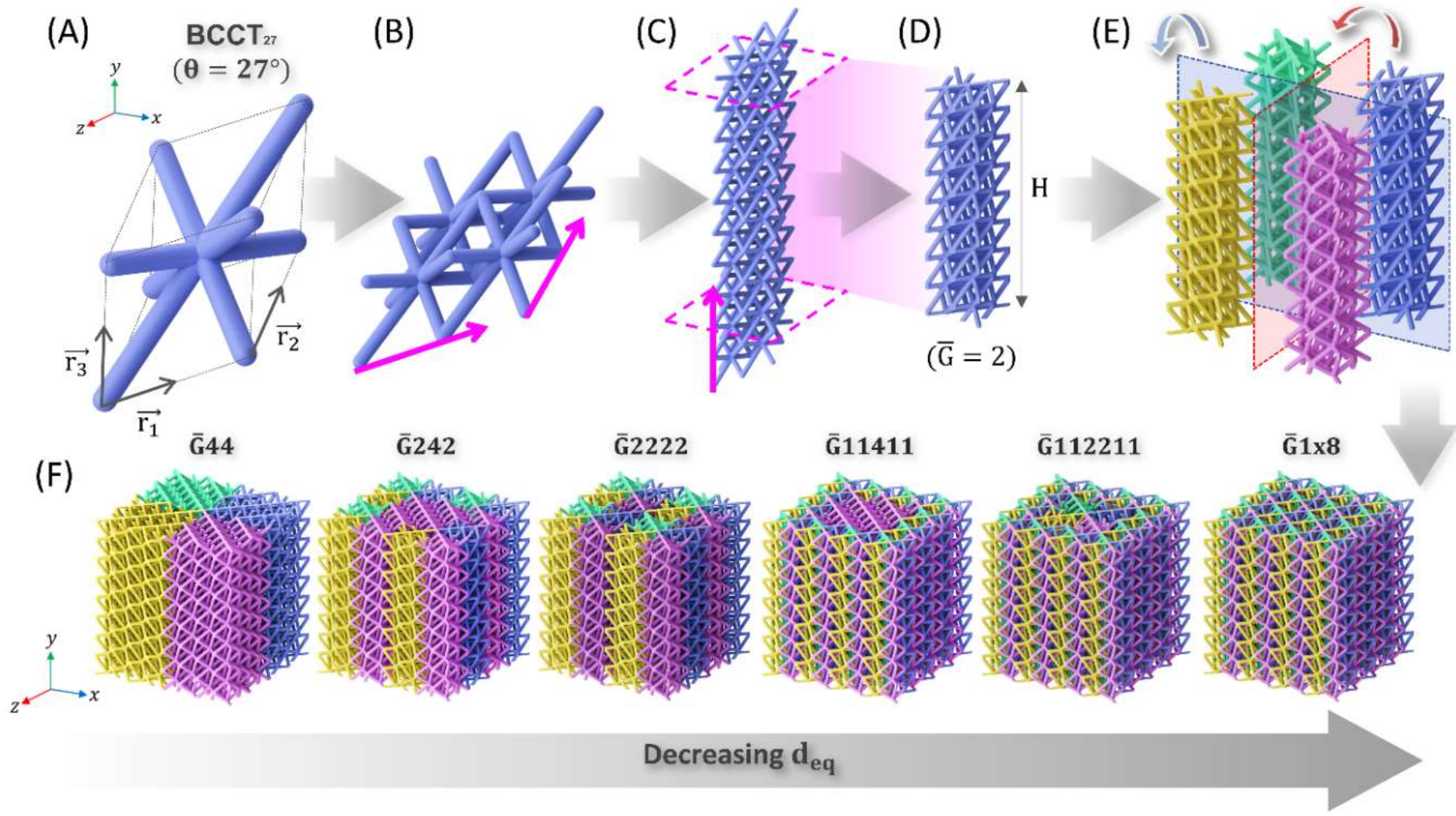

**S-8**: Design process for BCCT lattices with varied meta-grain size (A) BCCT27 cell (B) tessellated along $\vec{r_1}$ and $\vec{r_2}$ vectors (C) tessellated along $\vec{r_3}$ vector and (D) trimmed to form meta-grain with $\bar{G} \times \bar{G} \times \bar{H}$ cells. (E) Meta-grain mirrored about XY and YZ planes to form meta-grains of different orientation. (F) Meta-grains arranged to form Meta-Grain (MG): $\bar{G}$44, $\bar{G}$2222 and $\bar{G}$1x8 lattices and Meta-Harmonic (MH): $\bar{G}$242, $\bar{G}$11411 and $\bar{G}$112211 lattices

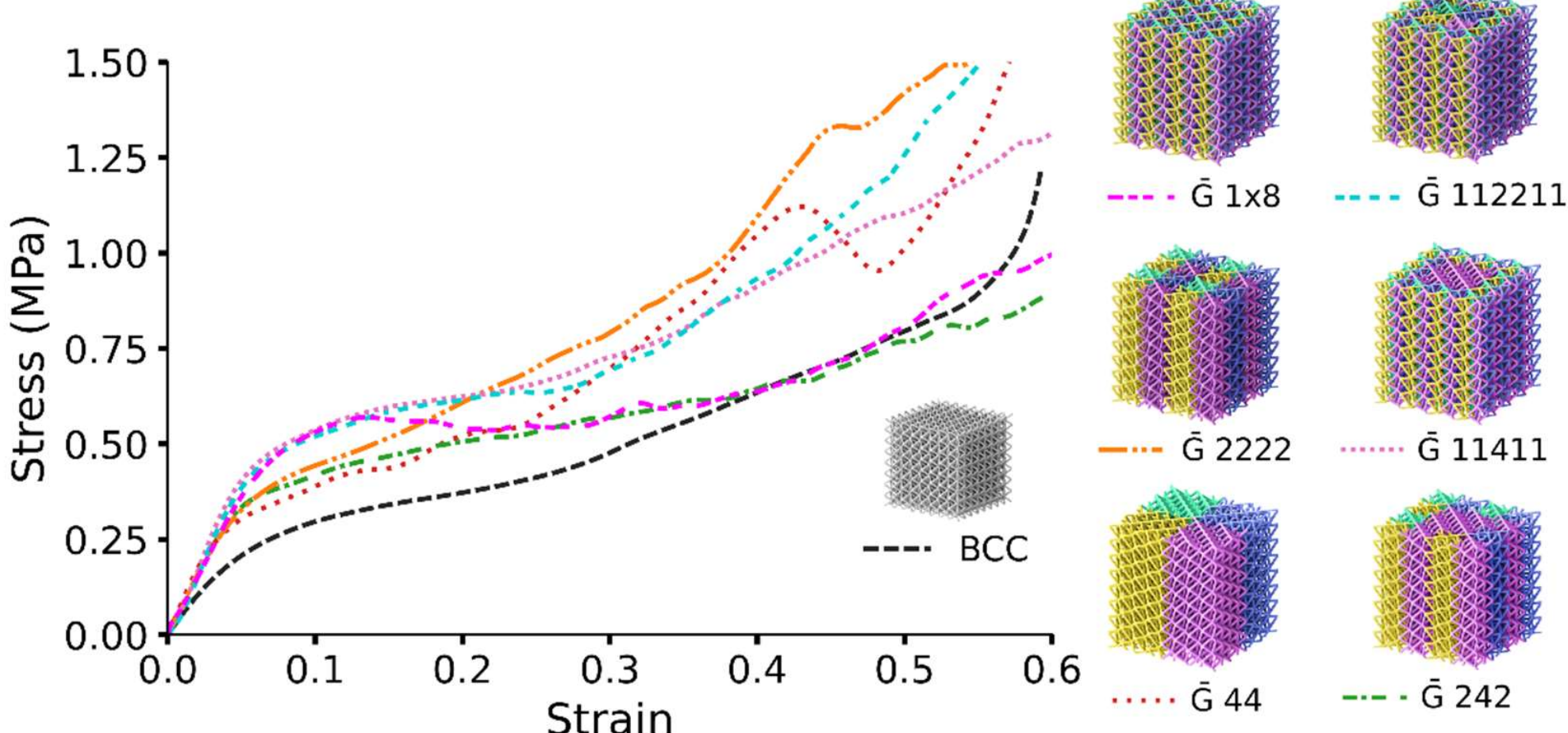

**S-9**: Stress-strain response of MG and MH BCCT$_{27}$ lattices ($\bar{\rho} = 0.14$) printed by SLA in Rigid 4K polymer and tested under uniaxial compression in an Instron testing machine

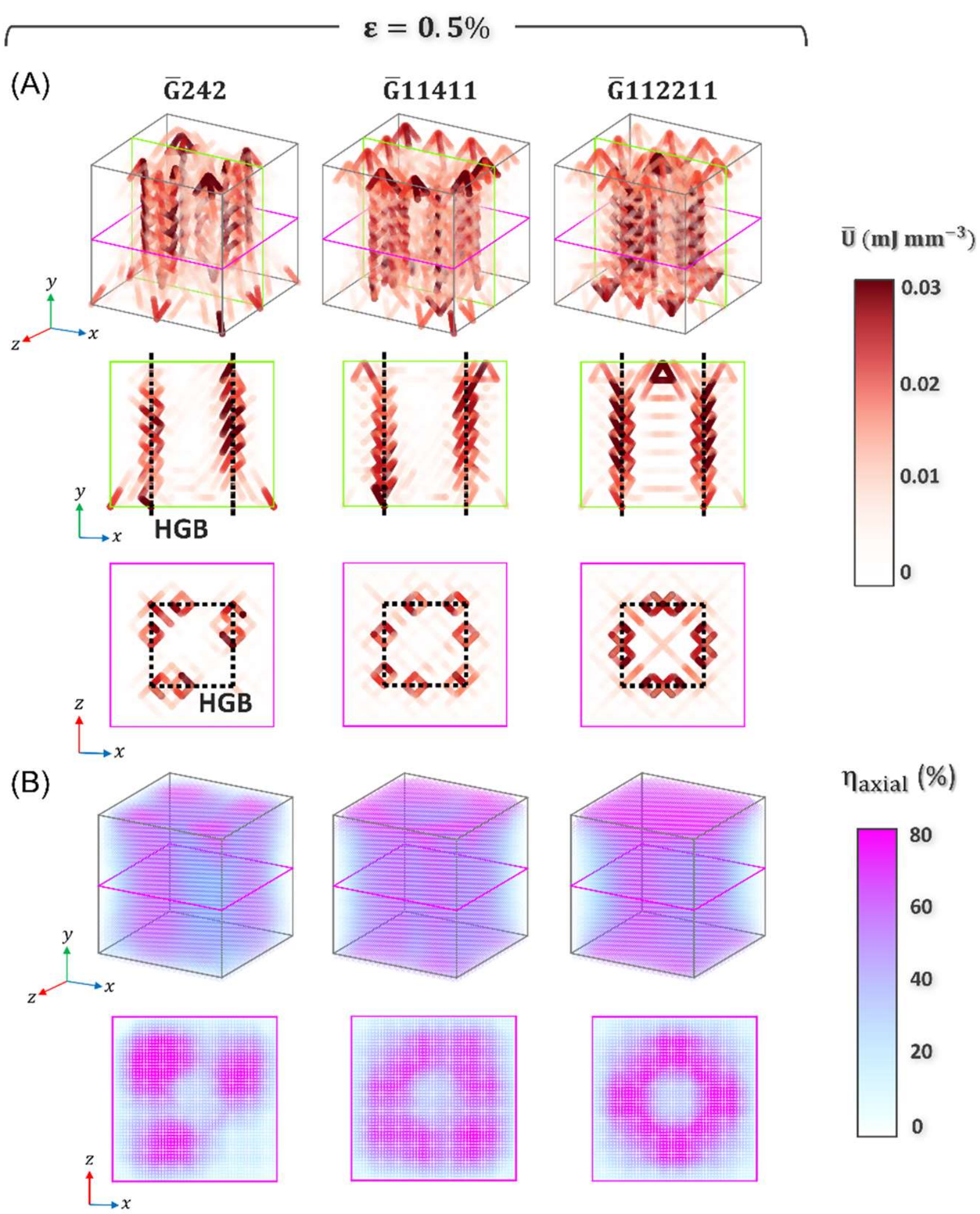

**S-10**: 3D and 2D section maps of (A) axial strain energy density and (B) axial proportion of strain energy, $\eta_{axial}$, for BCCT27 ($\bar{\rho} = 0.14$) meta-grain and meta-harmonic lattices at 0.5% global strain. There is localisation of axial energy around harmonic grain boundaries; grain boundaries between meta-grains of differing size. Non-symmetrical distributions of axial energy in the $\bar{G}242$ and $\bar{G}11411$ lattices results from non-symmetrical lattice geometry

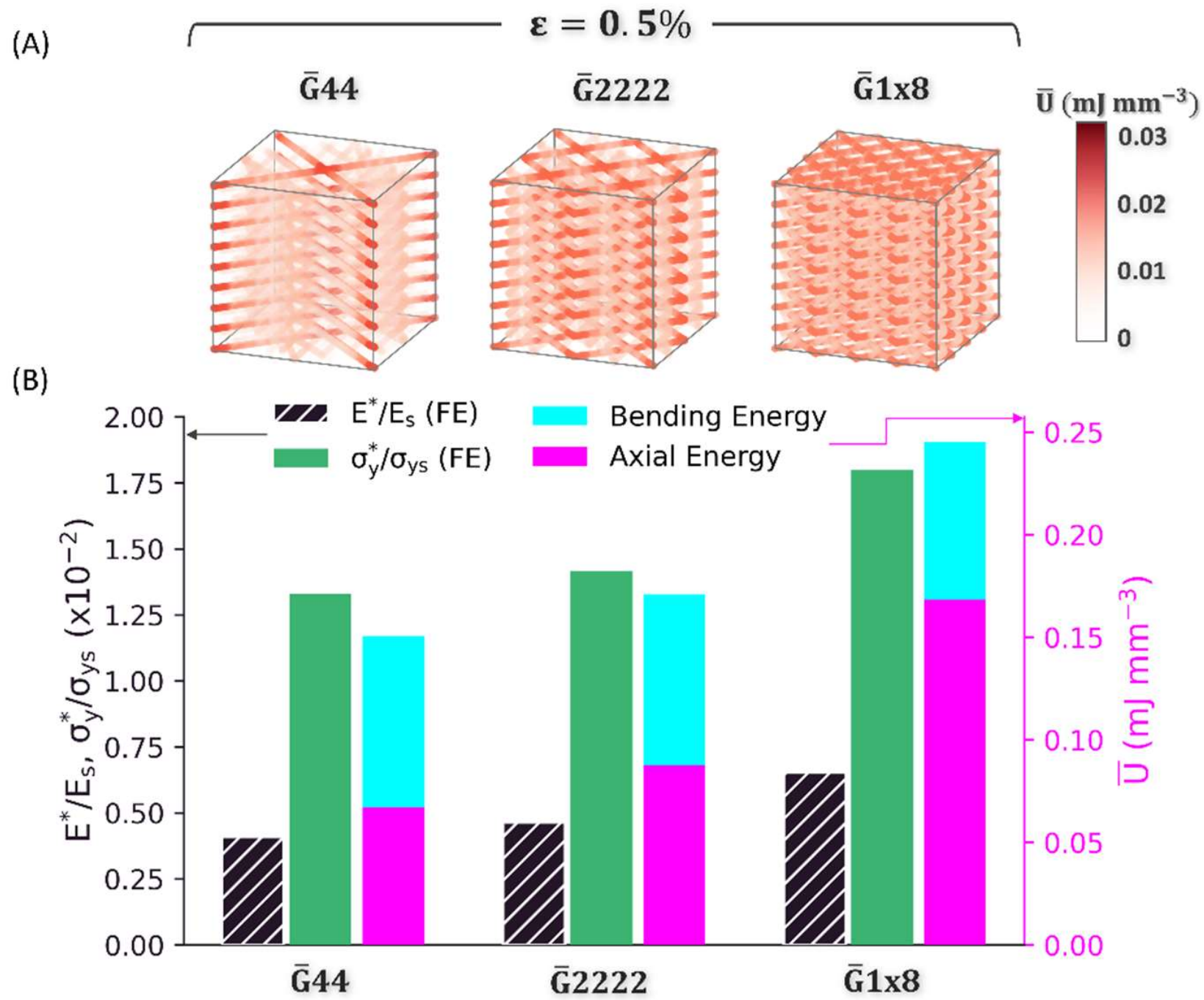

**S-11**: (A) 3D maps of axial strain energy plotted for UCs of Meta-Grain (MG) lattices $\bar{G}44$, $\bar{G}2222$ and $\bar{G}1x8$. Note that results have been plotted for UCs of size 8d in all orthogonal directions for comparison. (B) Relative stiffness, strength, bending strain energy and axial strain energy increase with decreasing meta-grain size for infinite-sized MG lattices.

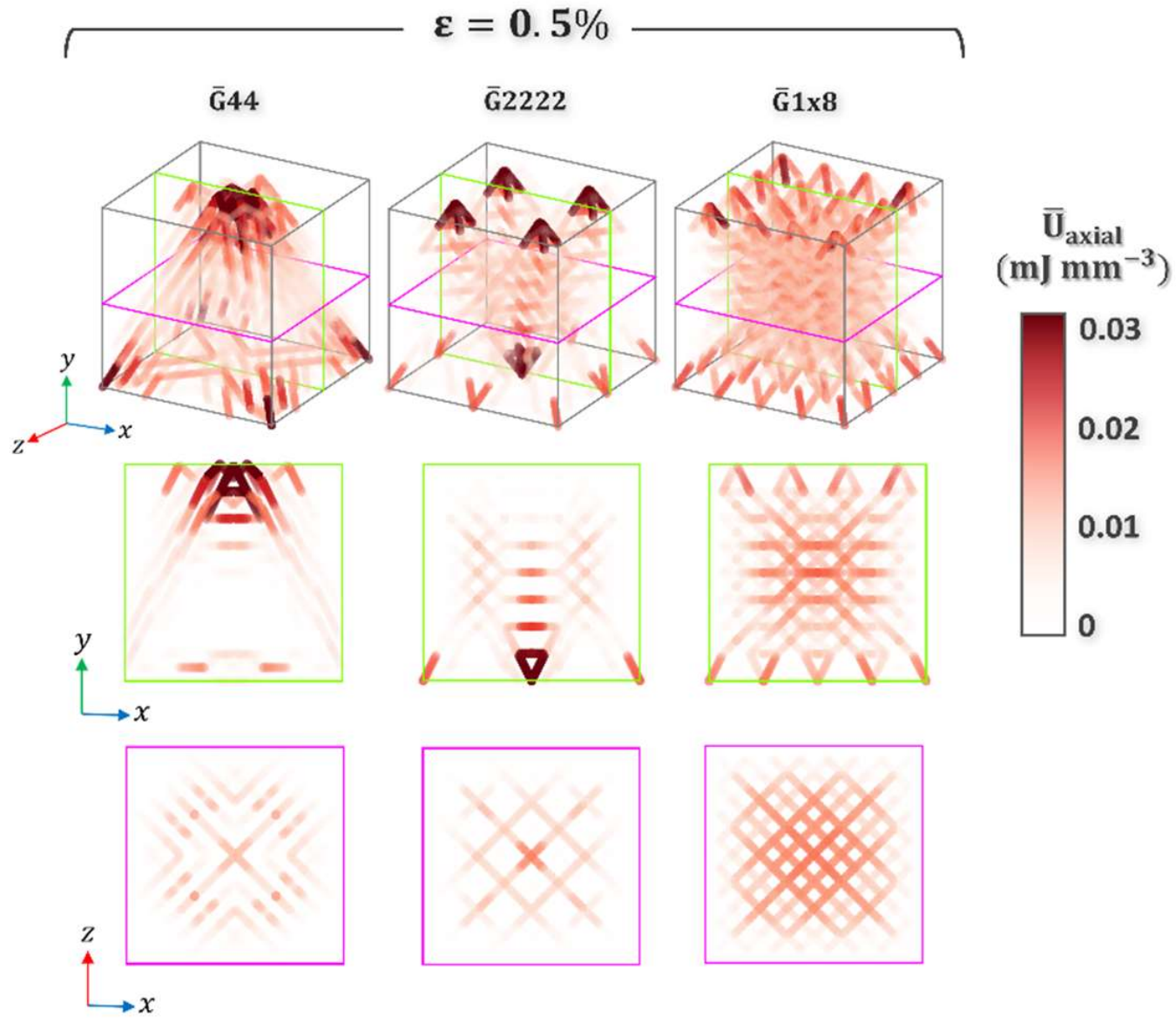

**S-12**: 3D and 2D section maps showing axial strain energy density for BCCT27 ($\bar{\rho} = 0.14$) meta-grain lattices at 0.5% global strain

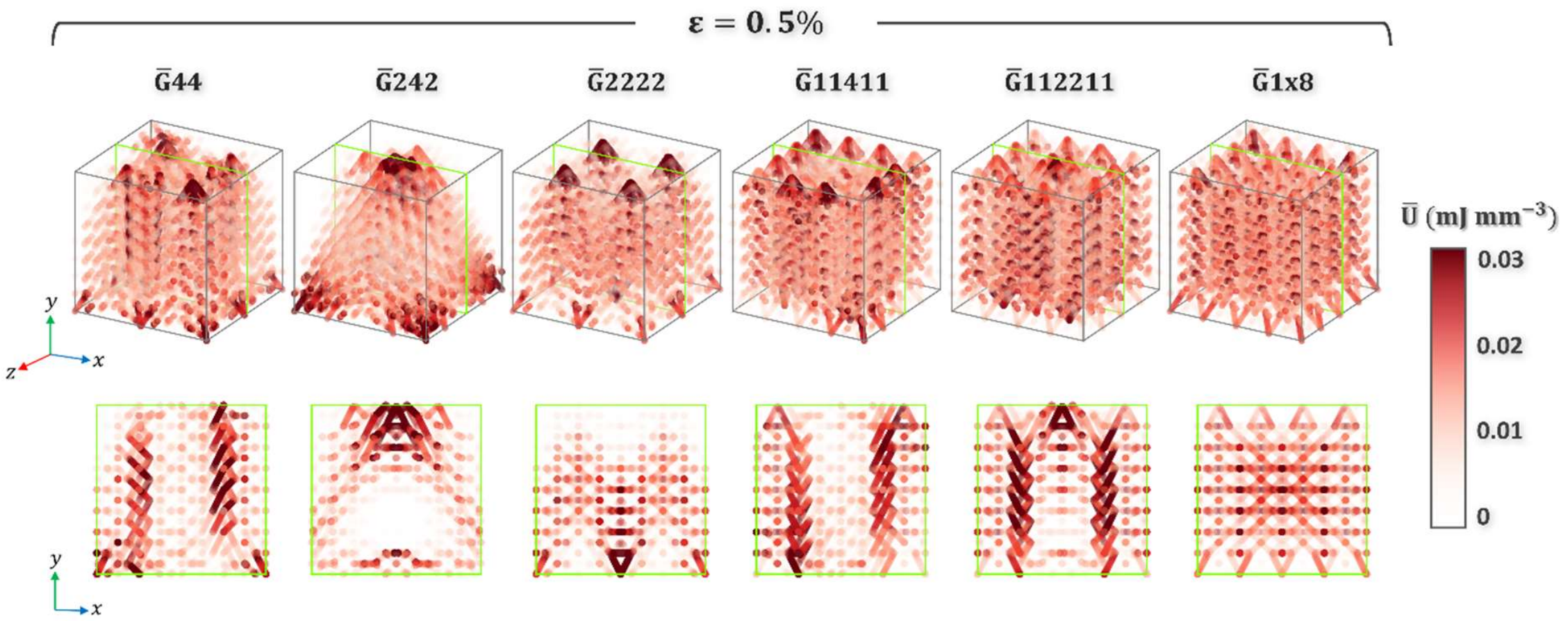

**S-13**: 3D and 2D section maps showing total strain energy density for BCCT27 ($\bar{\rho} = 0.14$) meta-grain and meta-harmonic lattices at 0.5% global strain

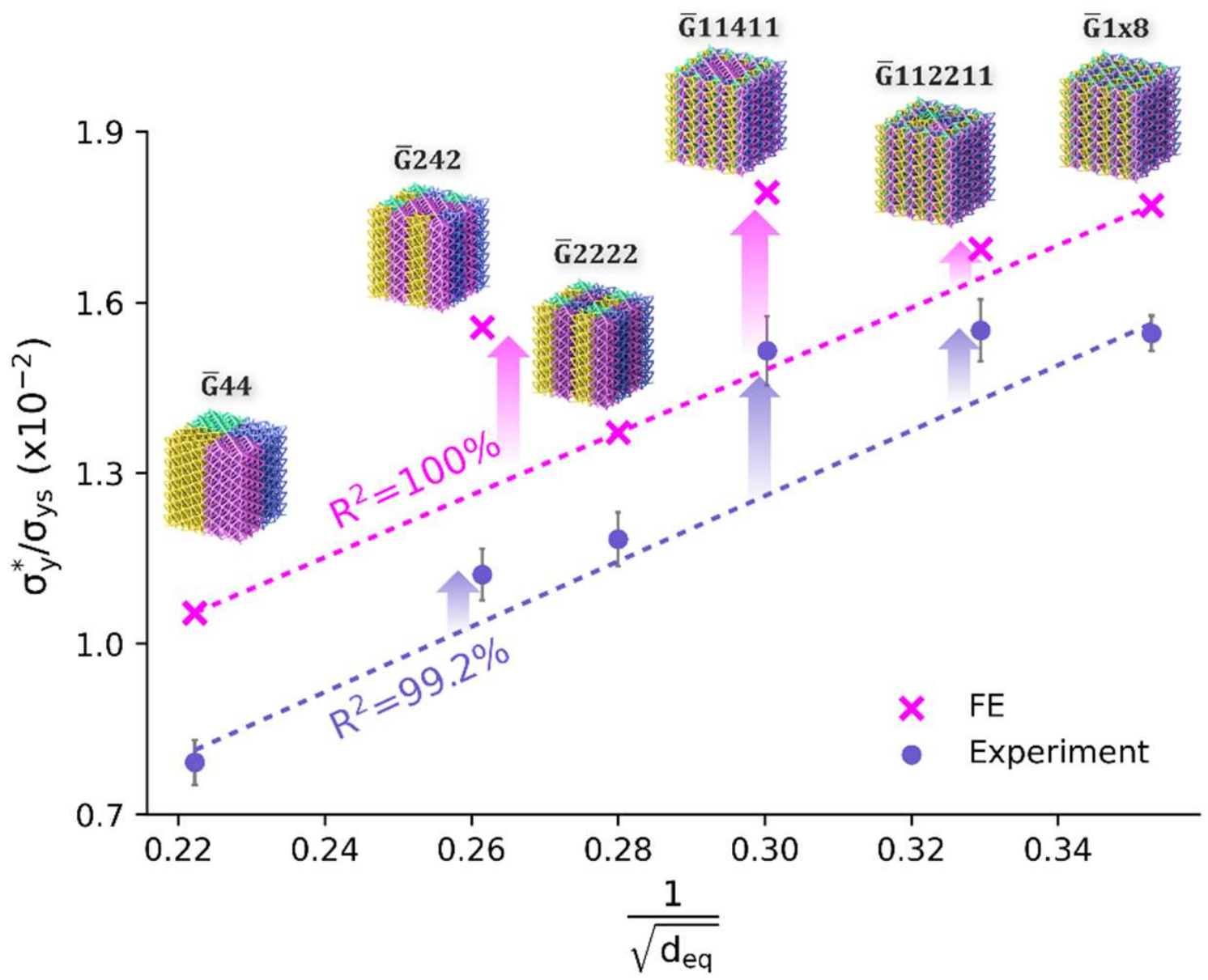

**S-14:** Relative yield strength plotted against $1/d_{eq}$ for BCCT27 ($\bar{\rho} = 0.14$) Meta-Grain (MG) and Meta-Harmonic (MH) lattices, determined both experimentally and through numerical simulation. The linear relationship for MG lattices agrees with the Hall-Petch relationship that describes grain boundary strengthening but the MH lattices outperform such predictions of strength.

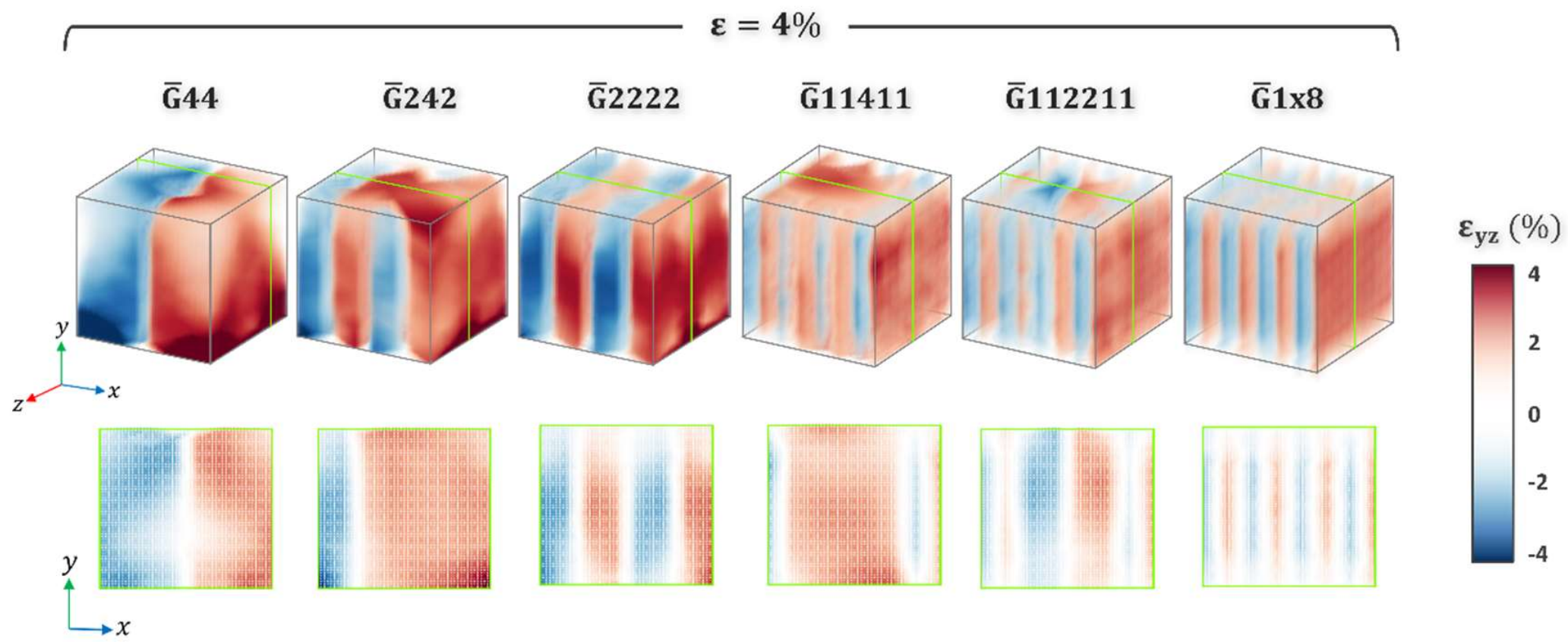

**S-15**: 3D and 2D section maps showing $\varepsilon_{yz}$ for BCCT27 ($\bar{\rho} = 0.14$) meta-grain and meta-harmonic lattices at 4% global strain

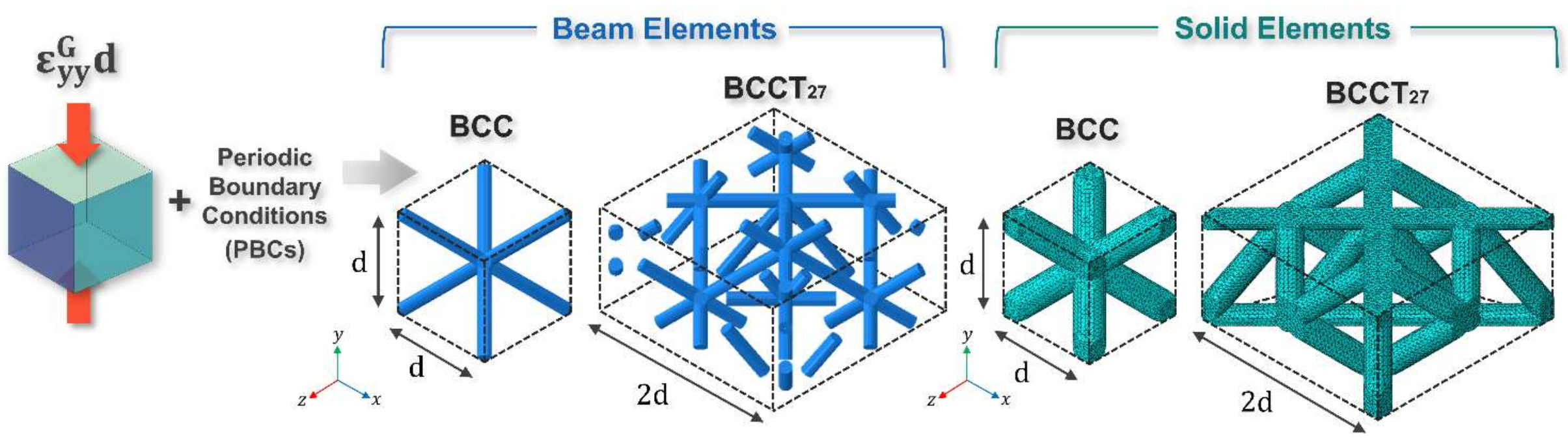

**S-16:** UC analysis set up as for BCC and BCCT27 lattices modelled with beam and solid elements. Periodic Boundary Conditions (PBCs) are applied to the element nodes located on the faces of the bounding box and a uniaxial displacement is applied to induce a uniaxial global strain ε

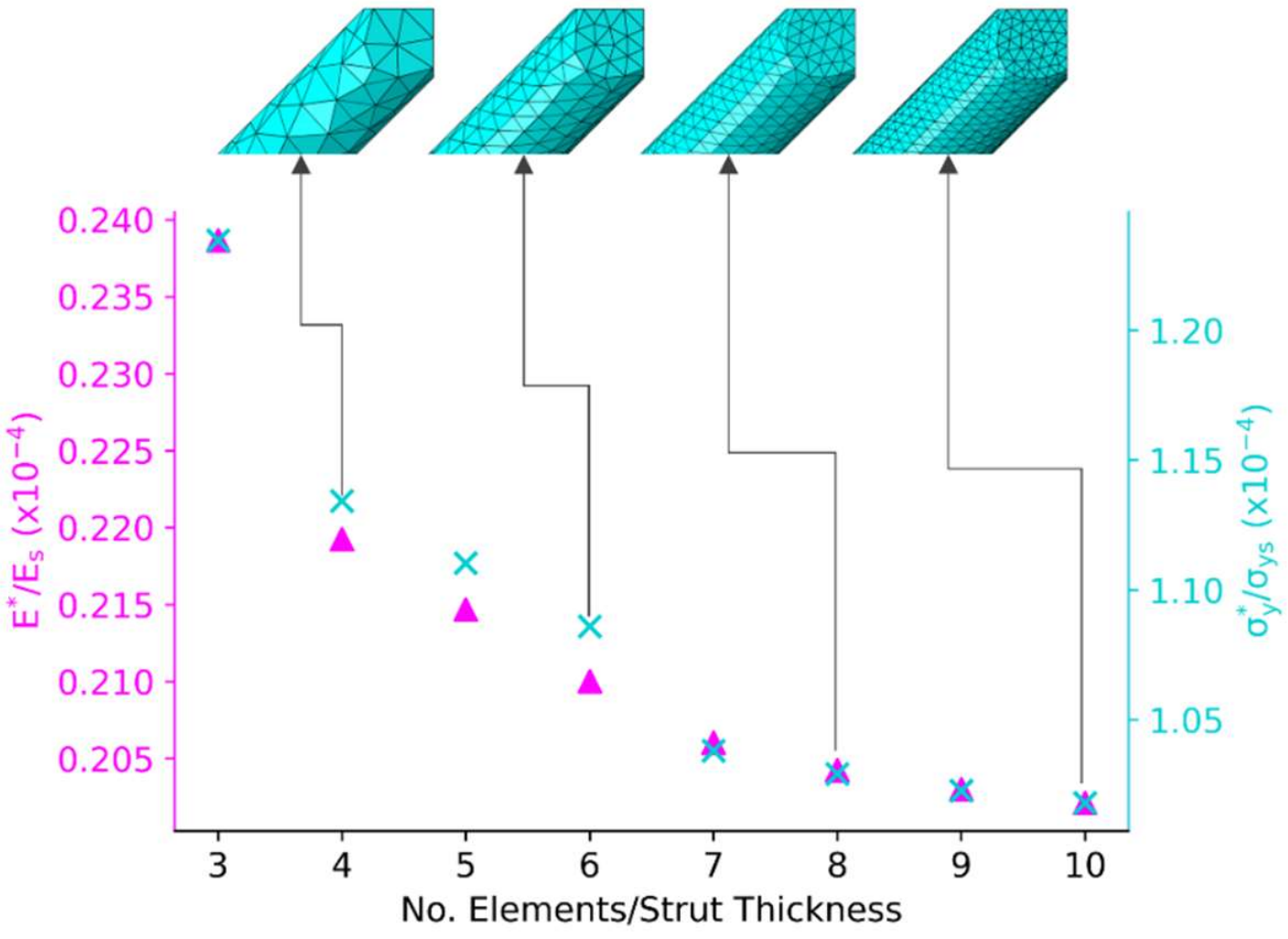

**S-17**: Mesh convergence for a BCC UC ($\bar{\rho} = 0.14$) modelled with C3D10 elements. $\frac{E^*}{E_s}$ and $\frac{\sigma_y^*}{\sigma_s}$ are plotted against number of elements across strut thickness and converge at 8 elements. Mesh convergence was repeated for beam element models and for BCCT27 lattices

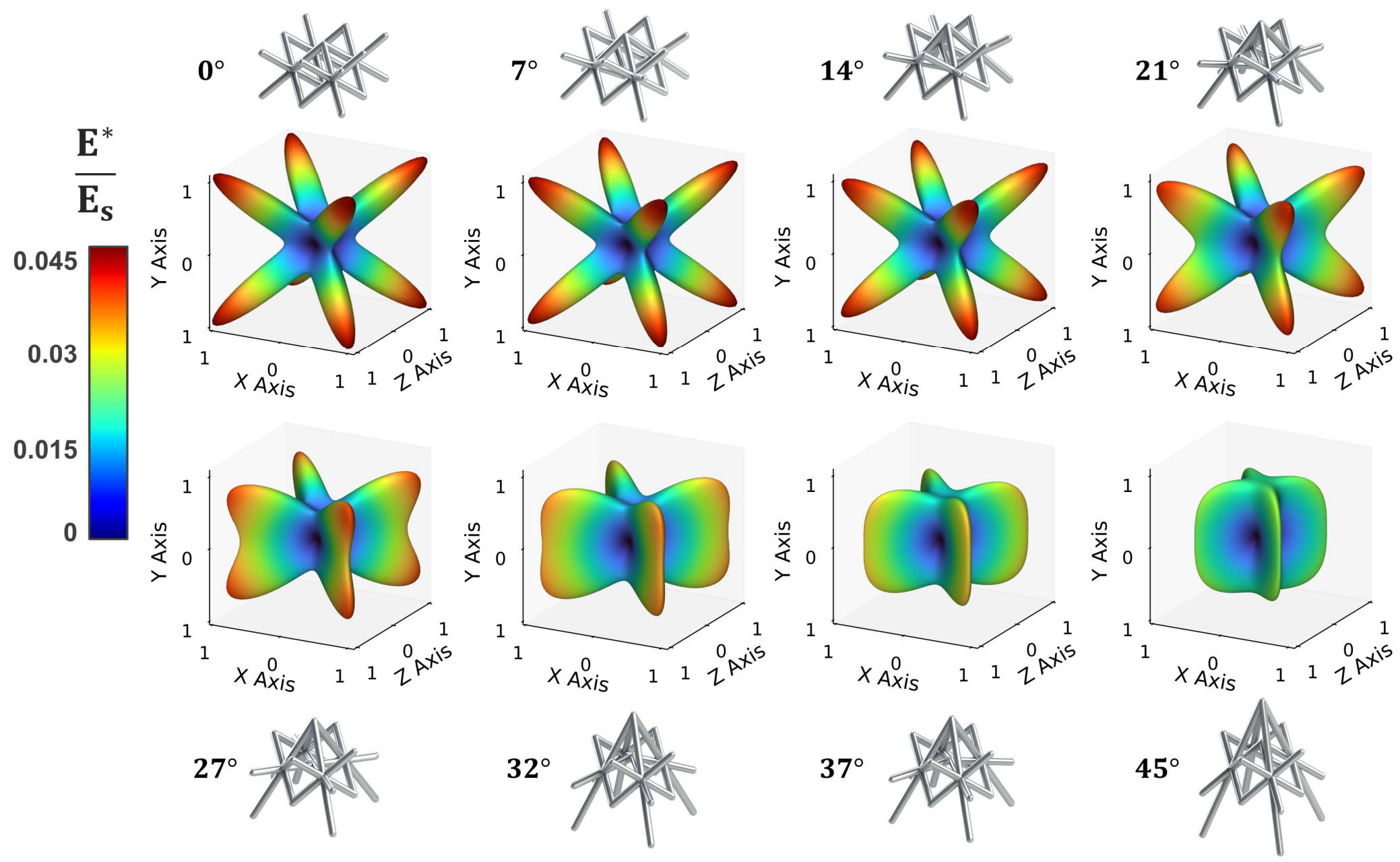

**S-18:** Surface plots of $E^*/E_s$ for BCCT lattices with $0° \leq \theta \leq 45°$ showing a monotonic decrease in anisotropy with increasing $\theta$.

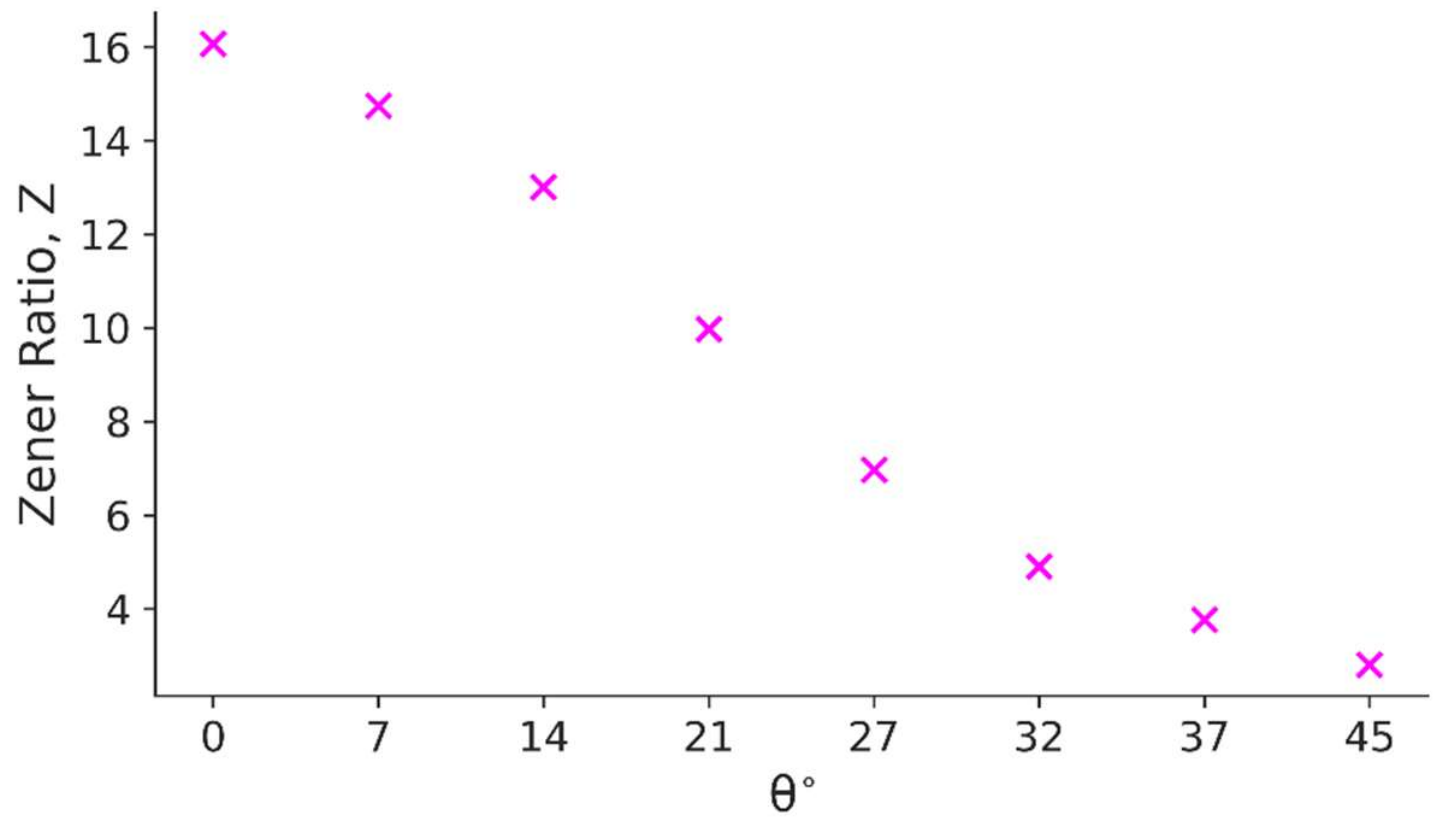

**S-19**: Zener ratio, Z plotted against $\theta$ for BCCT lattices with $0° \leq \theta \leq 45°$. For the BCC lattice, with $\bar{\rho} = 0.14$, $Z = 16.03$.

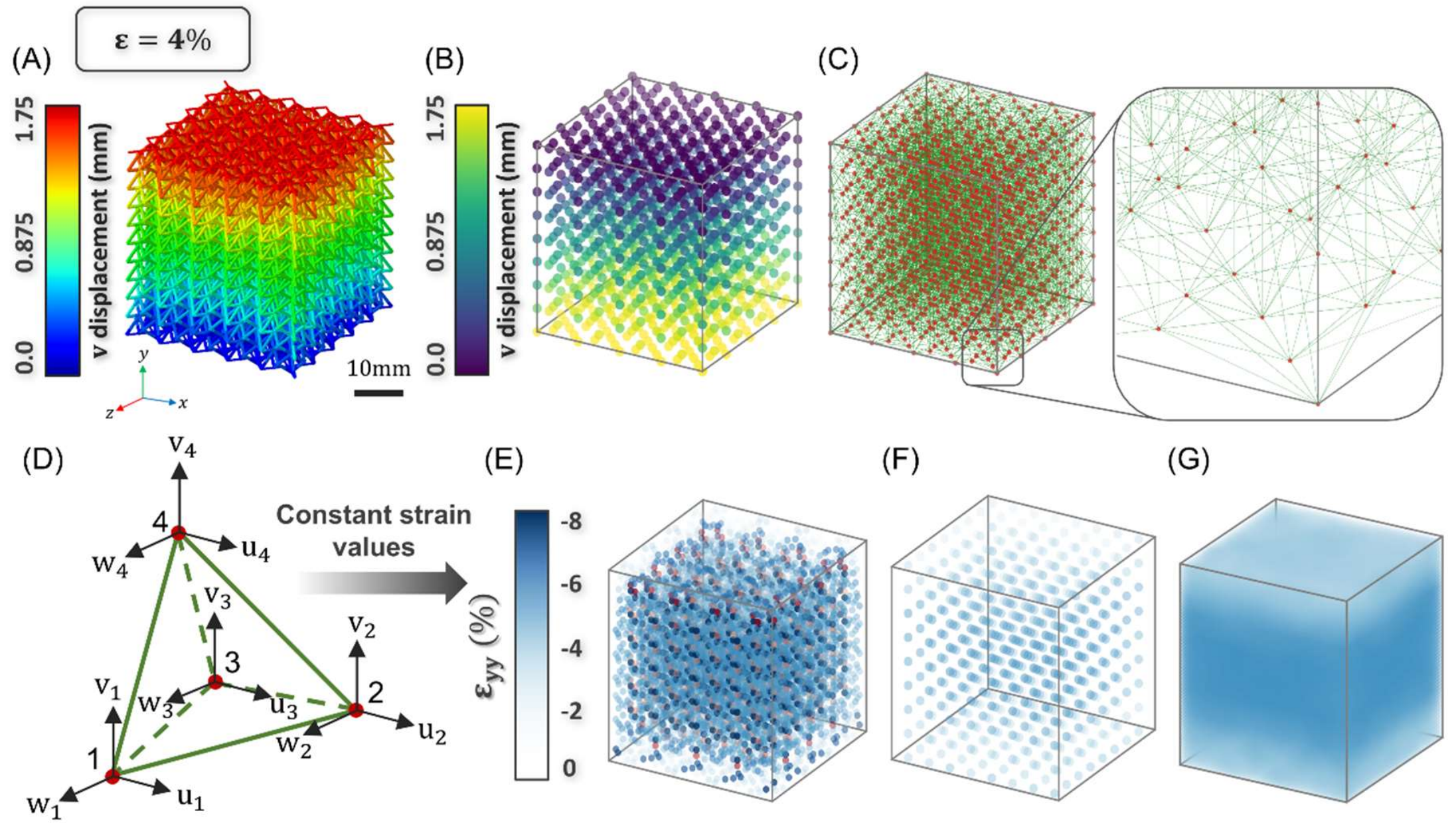

**S-20**: (A) Displacements output from Abaqus for element nodes (v displacement shown) (B) Displacements isolated for lattice nodes (v displacement shown) (C) 3D Delauney triangulation performed using SciPy package to separate lattice space into tetrahedral mesh (D) Typical tetrahedral element showing displacements $u_i, v_i$ and $w_i$ in the x, y and z directions, respectively for the $i^{th}$ node (E) Constant strain values calculated for each tetrahedral element and plotted in 3D at the element centres (F) Volume-weighted mean strain value calculated for each lattice cell and plotted in 3D at the cell centres (G) Strain values for a $60 \times 60 \times 60$ linearly spaced point cloud determined through 3D linear interpolation plotted in 3D ($\varepsilon_{yy}$ values shown)

## Tables S1 to S2

**Supplementary Table 1**: Scaling laws predicted by the G-A model for various relative properties of lattices

| Relative Property | Scaling Law | Ideal Exponent | Behaviour |
|---|---|---|---|
| Modulus, $\overline{E} = \frac{E^*}{E_s}$ | $\overline{E} = A\overline{\rho}^a$ | $a = 2$<br>$a = 1$ | Bending-dominated<br>Stretch-dominated |
| Yield strength, $\overline{\sigma}_y = \frac{\sigma^*}{\sigma_s}$ | $\overline{\sigma}_y = B\overline{\rho}^b$ | $\frac{3}{2} \leq b \leq 2$<br>$b = 1$ | Bending-dominated<br>Stretch-dominated |
| Elastic buckling strength, $\overline{\sigma}_{el} = \frac{\sigma^*_{el}}{E_s}$ | $\overline{\sigma}_{el} = C\overline{\rho}^c$ | $c = 2$ | Buckling-dominated |

**Supplementary Table 2:** Pre-exponent and exponent values for BCC and BCCT lattices determined through UC analysis

| | | **BCC** | | **BCCT$_{32}$** | | **BCCT$_{37}$** | | **BCCT$_{45}$** | |
|---|---|---|---|---|---|---|---|---|---|
| | | Value | $R^2$ | Value | $R^2$ | Value | $R^2$ | Value | $R^2$ |
| $\frac{E^*_{22}}{E_s} = A\bar{\rho}^a$ | A | 0.15 | 1.00 | 0.08 | 1.00 | 0.10 | 1.00 | 0.11 | 1.00 |
| | a | **2.05** | | **1.08** | | **1.07** | | **1.06** | |
| $\frac{\sigma^*_{Y,22}}{\sigma_{Y,s}} = B\bar{\rho}^b$ | B | 0.28 | 1.00 | 0.15 | 1.00 | 0.26 | 1.00 | - | - |
| | b | **1.81** | | **0.98** | | **1.08** | | **-** | |
| $\frac{\bar{\sigma}^*_{el}}{E_s} = C\bar{\rho}^c$ | C | - | - | 0.06 | 1.00 | 0.07 | 1.00 | - | - |
| | c | **-** | | **2.06** | | **2.08** | | **-** | |